%% file: main.tex
\documentclass[sigconf]{acmart}

\author{Xuanqi Gao}
\affiliation{
  \institution{Xi'an Jiaotong University}
  \city{Xi'an}
  \country{China}
}
\email{gxq2000@stu.xjtu.edu.cn}

\author{Weipeng Jiang}
\affiliation{
  \institution{Xi'an Jiaotong University}
  \city{Xi'an}
  \country{China}
}
\email{lenijwp@stu.xjtu.edu.cn}

\author{Juan Zhai}
\affiliation{
  \institution{University of Massachusetts}
  \country{Amherst, MA, USA}
}
\email{juanzhai@umass.edu}

\author{Shiqing Ma}
\affiliation{
  \institution{University of Massachusetts}
  \country{Amherst, MA, USA}
}
\email{shiqingma@umass.edu}

\author{Xiaoyu Zhang}
\affiliation{
  \institution{Xi'an Jiaotong University}
  \city{Xi'an}
  \country{China}
}
\email{zxy0927@stu.xjtu.edu.cn}

\author{Chao Shen}
\affiliation{
  \institution{Xi'an Jiaotong University}
  \city{Xi'an}
  \country{China}
}
\email{chaoshen@mail.xjtu.edu.cn}

\input{pkgs}

\AtBeginDocument{%
  \providecommand\BibTeX{{%
    \normalfont B\kern-0.5em{\scshape i\kern-0.25em b}\kern-0.8em\TeX}}}

\setcopyright{none}
\renewcommand\footnotetextcopyrightpermission[1]{} 

\begin{document}

\newcommand{\sys}{\mbox{\textsc{Ballot}}\xspace}
\newcommand{\update}[1]{\textcolor{blue}{#1}}

\title{Efficient DNN-Powered Software with Fair Sparse Models}

\input{abstract.tex}


\maketitle

\input{body/Introduction.tex}
\input{body/Background.tex}

\input{body/Design.tex}
\input{body/Evaluation.tex}
\input{body/RelatedWork.tex}
\input{body/Conclusion.tex}

\newpage
\bibliographystyle{ACM-Reference-Format}
\bibliography{BALLOT}

\end{document}

%% file: pkgs.tex
\usepackage{tikz}
\usepackage{amsmath}
\usepackage{filecontents}
\usepackage[flushleft]{threeparttable}
\usepackage{amsmath,amsfonts}
\usepackage{graphicx}
\usepackage{textcomp}

\usepackage[normalem]{ulem}

 \usepackage{url}

\usepackage{lipsum,tabularx}

\usepackage{multicol}
\usepackage{multirow}

\usepackage{colortbl}
\usepackage{booktabs}
\usepackage{setspace}

\hypersetup{
  linkcolor={blue!70!black},
  citecolor={red!70!black},
  urlcolor={blue!70!black}
}
\def\Snospace~{\S{}}

\usepackage[T1]{fontenc}

\usepackage{algorithm}
\usepackage{algorithmicx}
\usepackage[noend]{algpseudocode}

\usepackage{balance}

\usepackage{bm}
\usepackage{fp}
\usepackage{siunitx}
\usepackage{amsthm}

\usepackage[labelfont=bf,font=small,skip=5pt]{caption}
\usepackage{subcaption}

\usepackage{comment}

\usepackage{fancyhdr}
\usepackage{framed}
\colorlet{shadecolor}{blue!20}

\usepackage{tikz}

\usepackage{xspace}

\newcommand{\boxbeg}{
  \vspace{2px}
  \noindent\begin{tabular}{|l|}\hline
    \begin{minipage}{3.2in}
      \vspace{2px}
      \noindent
      }

      \newcommand{\boxend}{
      \vspace{2px}
    \end{minipage} \\ \hline
  \end{tabular}
  \vspace{-10pt}
}

%% file: abstract.tex
\begin{abstract}
    With the emergence of the Software 3.0 era, there is a growing trend of compressing and integrating large models into software systems, with significant societal implications. 
    Regrettably, in numerous instances, model compression techniques impact the fairness performance of these models and thus the ethical behavior of DNN-powered software. 
    One of the most notable example is the Lottery Ticket Hypothesis~(LTH), a prevailing model pruning approach.
    This paper demonstrates that fairness issue of LTH-based pruning arises from both its subnetwork selection and training procedures, highlighting the inadequacy of existing remedies.
    To address this, we propose a novel pruning framework, \sys, which employs a novel conflict-detection-based subnetwork selection to find accurate and fair subnetworks, coupled with a refined training process to attain a high-performance model, thereby  improving the fairness of DNN-powered software.
    By means of this procedure, \sys improves the fairness of pruning by 38.00\%, 33.91\%, 17.96\%, and 35.82\% compared to state-of-the-art baselines, namely Magnitude Pruning, Standard LTH, SafeCompress, and FairScratch respectively, based on our evaluation of five popular datasets and three widely used models.
    Our code is available at \url{https://anonymous.4open.science/r/Ballot-506E}.
\end{abstract}


%% file: body/Introduction.tex
\section{Introduction}\label{sec:intro}

We envisage that the Software 3.0 era, characterized by software powered by large-scale models, will pave the way for numerous potential applications, such as artificial intelligence generated content~(AIGC) and autonomous driving, which are poised to exert substantial influence on societal transformations~\cite{shahNvidiaCEOHuang2023,zhaoSurveyLargeLanguage2023}. 
The magnitude of AI software has notably surged, primarily driven by the escalating size of deep neural network models~\cite{mishraSurveyDeepNeural2020}. 
For instance, state-of-the-art computer vision models now encompass over 15 billion parameters, while large language models like GPT-3 exceed 175 billion parameters, demanding nearly 1TB of storage exclusively for the model itself~\cite{brownLanguageModelsAre2020}. 

However, the deployment of large-scale AI software introduces several challenges.
These large models require substantial memory and storage resources during deployment, presenting difficulties in running them on resource-constrained devices like edge devices, smartphones, or wearables~\cite{han2015deepcom}. 
Furthermore, the size of these models often leads to slow inference times due to the sheer number of parameters and computations involved, which can be particularly problematic in time-critical applications such as real-time object detection or autonomous driving~\cite{liang2021pruning}.
Additionally, when AI models are deployed on cloud servers and accessed remotely by client devices, the transfer of large model files over limited bandwidth can result in high latency and increased data consumption~\cite{wareHowGiantAI2022}.
In addition to these deployment difficulties, the resource-intensive nature of large AI applications poses further challenges. 
The computational demands of these models significantly consume energy, which becomes a critical concern for battery-operated devices or data centers striving for energy efficiency~\cite{luccioni2022estimating}. 
Moreover, the carbon footprint associated with the training and deployment of large models can be substantial, contributing to environmental concerns~\cite{schwartz2020green}.

To address the aforementioned challenges, researchers employ model compression techniques, which aim to reduce the size and complexity of AI models while retaining their performance.
Various model compression techniques have been introduced, encompassing model pruning\cite{polyak2015channel,luo2017entropy,he2019filter,li2016pruning,frankleLotteryTicketHypothesis2018}, model quantization\cite{ni2020wrapnet,jacob2018quantization}, and knowledge distillation\cite{hinton2015distilling,ba2014deep,bengio2013representation}. 
Notably, the model pruning algorithm stands out as one of the most widely adopted methods in this domain\cite{mishraSurveyDeepNeural2020,liuOnDemandDeepModel2018}.
Pruning involves the selective removal of unnecessary parameters, connections, or entire neurons, leading to a more streamlined and efficient model.
As a state-of-the-art method, the Lottery Ticket Hypothesis~(LTH) has garnered significant attention and extensive research efforts~\cite{frankleLotteryTicketHypothesis2018}. 
The hypothesis posits the existence of a winning ticket, namely a properly pruned subnetwork combined with the original weight initialization, which can achieve competitive performance comparable to that of the original dense network. 
This discovery underscores the immense potential for efficient training and network design in the realm of deep learning.

Unfortunately, among the various research on the LTH and also our experiment results~(see \autoref{sec:lth}), LTH-based pruning methods suffer from model bias problem~\cite{maSanityChecksLottery2021}.
Through our analysis, we found that the bias problem is caused by both ticket selection and ticket training.
The process of ticket selection entails the elimination of redundant neurons, resembling high-dimensional feature selection for images, and potentially introducing bias. 
Fairness and accuracy can exhibit conflicts during ticket selection, with prevailing LTH-based pruning methods prioritizing accuracy, resulting in tickets that lack fairness.
Even after obtaining a ticket that balances accuracy and fairness, there is no assurance that the ticket can be trained to reach its theoretical performance upper limit. 
The training process faces potential issues of overfitting and underfitting, impacting the model's ability to generalize and make accurate and fair predictions.
Current efforts to enhance the LTH-based pruning primarily concentrate on improving its efficiency and accuracy performance, neglecting its fairness concerns.

Inspired by ethics-aware software engineering~\cite{aydemirRoadmapEthicsawareSoftware2018,brunSoftwareFairness2018,galhotraFairnessTestingTesting2017}, we present \sys, a novel fairness-aware pruning framework by revised winning ticket finding and training for deep neural networks.
The key idea of \sys is that it follows the best practice of software engineering by first observing and analyzing the root cause of the bias problem.
Concretely, it performs a conflict detection between the fairness optimization direction and the accuracy optimization direction of the model during the model training process to identify neurons with the highest degree of conflict.
These neurons are more likely to induce the model to optimize towards greater variance among distinct subgroups during training,
leading to biased model behavior and subsequently giving rise to instances of discriminatory behavior in the software.
Each neuron is assigned a score representing its importance in training optimization conflicts, and a corresponding mask is generated to remove these neurons, resulting in the acquisition of an accurate and fair ticket.
To ensure that the ticket can eventually be trained into high-performance models, we also refine the training procedure.
This involves adapting the learning rate based on the training condition to facilitate the model's learning of richer features.
Furthermore, we verify the fairness performance of the model upon the completion of training. 
If the fairness performance is inferior to that of the original model, we reload the weights from earlier training rounds~(the specific number of rounds predetermined by empirical considerations) and iteratively retrain the model until further performance improvement is unattainable.
This refined training approach is designed to maximize the ticket's performance potential.

\sys has been implemented as a self-contained toolkit.
Our experiments on CIFAR-100, TinyImageNet, and CelebA datasets show that \sys can effectively mitigate the pruning fairness problem while maintaining accuracy, compared with existing pruning methods~\cite{han2015learning,frankle2018lottery,zhu2022safety,tangFairScratchTickets2023}.


Our contribution can be summarized as follows:

\begin{itemize}
    \item We propose a novel fairness-aware DNN pruning framework. It leverages conflict-detection-based mask generation to find fair and accurate tickets, and training refinement to achieve optimal performance.
    \item We develop a prototype \sys based on the proposed idea, and evaluate it with CIFAR-100, TinyImageNet and CelebA.
    On average, \sys improves fairness by 38.00\%, 33.91\%, 17.96\%, and 35.82\% compared to state-of-the-art baselines, namely Magnitude Pruning, Standard LTH, SafeCompress, and FairScratch, outperforming all baselines in terms of fairness.
    \item Our implementation, configurations and collected datasets are available at~\cite{AnonymizedRepository}.
\end{itemize}

%% file: body/Background.tex
\section{Background and Motivation}\label{sec:bg}

\subsection{DNN Software and DNN Model Deployment}


The continuous advancements in deep neural network~(DNN) research and the availability of powerful computing resources have contributed to the widespread adoption of DNNs in various applications, making DNN-powered software a key driver of innovation in the field of software engineering~\cite{martinez-fernandezSoftwareEngineeringAIBased2022}.
DNN-powered software refers to applications and systems that leverage the capabilities of DNNs for various tasks, which are often developed with deep learning frameworks and libraries such as PyTorch~\cite{TorchNnPyTorch}, TensorFlow~\cite{abadiTensorFlowSystemLargeScale2016},  and Keras~\cite{gulli2017deepkeras}.
However, as the capabilities of deep learning models gradually increase, their scale also becomes larger and larger, naturally taking up a lot of storage space~\cite{nanDeepModelCompression2019}.
The deployment DNN software encounters complexities due to the diverse software landscape and deployment demands on various mobile platforms. 
Confronted with resource limitations, such as computational power, storage capacity, and battery life, especially in embedded devices like mobile devices, the imperative arises for mobile models to adhere to stringent conditions encompassing reduced model size, diminished computational complexity, and minimal battery power consumption~\cite{liuOnDemandDeepModel2018}. 
Consequently, a challenge is raised for DNN-powered software developers: how can a DNN-powered software be effectively deployed to meet performance requisites on platforms characterized by constrained resources? 
The compression of DNN models emerges as an indispensable operation in the deployment of DNN-powered software.

Numerous studies emphasize the requisite step of model compression for deploying DNN models across diverse platforms~\cite{nanDeepModelCompression2019,liuOnDemandDeepModel2018}. 
Predominantly employed techniques for DNN model compression encompass model quantization~\cite{ni2020wrapnet,tailor2020degree}, model pruning~\cite{polyak2015channel,luo2017entropy}, and knowledge distillation~\cite{hinton2015distilling,bengio2013representation}. 
Among these, the model pruning is widely favored and commonly utilized, due to its capability to directly reduce the number of computational operations involved, fundamentally alleviating computation and memory pressures~\cite{li2023canpruning,liang2021pruning}.
Popular AI software development frameworks such as PyTorch~\cite{pytorchpruning}, Tensorflow~\cite{tensorflowpruning} and PaddlePaddle~\cite{paddlepdpruning}, all integrate pruning-based model compression toolkits.

Model pruning is a kind of a prominent and effective compression technique in the field of large-scale AI software deployment.
The motivation behind model pruning stems from the recognition that many deep learning models, especially those with an over-parameterized nature, contain numerous redundant or less critical parameters.
By selectively removing unimportant components or parameters from a model, this approach achieves a remarkable reduction in model size, rendering it more lightweight and computationally efficient as well as not significantly sacrificing model performance.
In a DNN, the architecture consists of layers of interconnected nodes~(neurons), and each connection between nodes is associated with a weight. 
These weights, along with biases, are the parameters of the model, and they are collectively represented by the symbol \(\theta\).
A DNN model \(f(\theta)\) is a function that takes input data and produces output based on these parameters.
More formally, model pruning aims to produce a lightweight model \(f(\theta^*)\) satisfying the following conditions:
\begin{equation}\label{eq:prunerate}
    \frac{|\theta^*|}{|\theta|} \leq \Omega
\end{equation}
\begin{equation}\label{eq:pruneperformance}
    Acc(f(\theta)) - Acc(f(\theta^{*})) \leq \epsilon
\end{equation}
The \(Acc(\cdot)\) measures accuracy, which is the most frequently utilized performance metric.
\(|\theta^*|\) and \(|\theta|\) are the numbers of parameters, \(\epsilon\) and \(\Omega\) are the tolerable decline of performance and the sparsity metric respectively.
Sparsity metric \(\Omega\) declares the level of pruning, for instance, the \(\Omega = 0.05\) requires the removal of 95\% of the parameters, i.e., the pruning rate with 0.95.
In practice, the goal is to maximize the accuracy of the pruned model, i.e. achieve a smaller \(\epsilon\) while adhering to a fixed sparsity metric \(\Omega\) to satisfy deployment requirements.
Pruned models can even achieve improved robustness and generalization, as the reduction of parameters helps alleviate issues such as overfitting~\cite{li2023canpruning,jin2022neural}.

\subsection{Fairness Problem in Pruning}
\label{sec:bgfairness}




While model pruning strikes a trade-off between accuracy and model size, recent research take attention to potential fairness concerns. 
For example,
Paganini points out that pruning should not only focus on overall performance metrics but should also consider performance differences across subgroups and individuals, i.e., the fairness of model~\cite{paganini2020prune}.
In machine learning, model fairness refers to the ethical and unbiased treatment of individuals or groups throughout the development, deployment, and utilization of machine learning models.
Consequently, the definition of fairness is also divided into two categories: individual fairness and group fairness.
Blakeney et al. find that the bias of the model grew progressively as the pruning rate increased~\cite{blakeney2021simon}.
For a pruning technique to be considered fair, the bias degree of the small model after pruning should not significantly increase compared to the metrics of the original model before pruning.
Formally, for an original model \(f(\theta)\) and a pruned model \(f(\theta^*)\), the fairness criterion can be defined as:
\begin{equation}\label{eq:prunefairness}
     d(f(\theta ^ *)) - d(f(\theta))  \leq \delta
\end{equation}
where \(d(\cdot)\) computes the bias degree (i.e. higher values represent lower levels of fairness), and \(\delta\) is the tolerance.
As previously mentioned, the definition of fairness is categorized into individual fairness and group fairness.
This paper specifically concentrates on group fairness. 
For datasets with sensitive features, fairness metrics such as demographic parity~(DP)~\cite{dworkFairnessAwareness2012} and equal opportunity~(EO)~\cite{hardtEqualityOpportunitySupervised2016} are commonly employed to assess whether the model exhibits discrimination against distinct groups. 
Since vision and natural language processing tasks usually have neither explicit sensitivity characteristics nor explicit favorable/unfavorable treatment,
we utilize class-wise variance~(CWV)~\cite{tianAnalysisApplicationsClasswise2021a} and maximum class-wise discrepancy~(MCD)~\cite{tianAnalysisApplicationsClasswise2021a} to measure variations in the model's performance across different groups, i.e. to measure the bias degree.

These metrics are formally defined below:

\smallskip
\noindent \textbf{Class-wise Variance~(CWV).}
Given a dataset with \(C\) classes and a given model, the accuracy on the subdataset belonging to \(c_{th}\) class is denoted as \(acc_{c}\).
\begin{equation}\label{eq:cwv}
    \text{CWV} = \frac{1}{C} \sum_{c=1}^C ( acc_c - \bar{acc} )^2,\; 
    \bar{acc}=\frac{1}{C}\sum_{c=1}^C acc_c
\vspace{-6pt}
\end{equation}

\smallskip
\noindent \textbf{Maximum Class-wise Discrepancy~(MCD).}
For a model, given the maximum and minimum class accuracy $acc^{+}$ and $acc^{-}$, MCD can be defined as:
\begin{equation}\label{eq:MCD}
    \text{MCD} = acc^{+} - acc^{-}
\vspace{-4pt}
\end{equation}

Intuitively, CWV represents the average discrepancy, while MCD assesses the extreme discrepancy among classes to estimate the fairness of a model. 
Larger values of these metrics indicate a higher degree of bias in the model. 
For example, for a software that includes a face recognition module, each class corresponds to one user, and if there are some users whose recognition accuracy is significantly lower than the average or other users (resulting in a large CWV or MCD), it implies unfair treatment, as users with equivalent qualifications receive disparate service quality.  
While our primary assessment of model fairness revolves around these two metrics, our approach readily extends to other fairness metrics such as DP and EO, contingent on the user defining advantageous outcomes based on the scenario.
We selected these metrics because they are well-established and commonly used metrics~\cite{benz2021robustness,gao2023fairCILciliate,tangFairScratchTickets2023} that are generally applicable to various tasks and datasets including both vision and natural language processing tasks discussed in this paper.

\subsection{Lottery Ticket Hypothesis}\label{sec:lth}

A recently proposed technique known as the Lottery Tickets Hypothesis~(LTH) has emerged as a rapidly growing method in the field of model pruning, which focuses on sparse trainable subnetworks within fully dense networks~\cite{frankleLotteryTicketHypothesis2018}.
The LTH posits the presence of a sparse subnetwork within a randomly initialized dense network, which can achieve test accuracy comparable to that of the original dense network within at most the same number of iterations during independent training.
This sparse subnetwork is known as the ``winning ticket''.
More specifically, for a randomly initialized dense neural network \(f(\theta_{0})\) parameterized by \(\theta_{0} \in \mathbb{R}^{|\theta_0|}\), LTH suggests searching a mask \(m \in (\{0,1\}^{|\theta_0|})\), i.e., the winning ticket is the sparse subnetwork \(f(\theta_{0} \odot  m)\), where \(\odot\) denotes the element-wise product and \(|\theta_0|\) denotes the numbers of parameters.
To find a winning ticket, it is common practice to first train the initialized network for some epochs and compute a mask based on the behaviors of the trained model.
LTH-based pruning usually follows the \textit{train-prune-retrain} pipeline as shown in \autoref{fig:overview}.
Given a dense neural network \(f(\theta_{0})\) initialized by the parameter \(\theta_{0} \in \mathbb{R}^{|\theta_0|}\) and the sparsity metric \(\Omega\), the process of finding a winning ticket and perform pruning can be expressed as: 
\begin{enumerate}
\item[i)] Train the network \(f(\theta_{0})\) for \(E\) epochs to get a well-trained model \(f(\theta)\).
\item[ii)] Remove parameters in \(\theta\) which satisfied the sparsity metric \(\Omega\),by creating a mask \(m \in (\{0,1\}^{|\theta_0|})\). Reset the remaining parameters to their initial values in \(\theta_{0}\), creating the winning ticket \(f(\theta_{0} \odot  m)\).
\item[iii)] Retrained the \(f(\theta_{0} \odot  m)\) to get final pruned model \(f(\theta^{*})\).
\end{enumerate}

\noindent Ideally, the performance of \(f(\theta ^ *)\) is expected to surpass that of \(f(\theta)\) within a limited number of training epochs, not exceeding \(E\). 
However, in a standard LTH setup~\cite{frankleLotteryTicketHypothesis2018}, a winning ticket can only be found when the learning rate is very small, and the performance advantage will become smaller when the learning rate is large.
How to find lotteries that actually win is still a problem worth exploring, but this does not detract from the practical value of LTH-based pruning to select high-value subnetworks.
A series of efforts were devoted to finding the most promising ticket, which can achieve higher performance\cite{ma2021sanity}.




\input{misc/LTHfair.tex}

Despite the remarkable success of LTH-based pruning in achieving high performance, it is crucial to recognize that it still grapples with the challenge of fairness. 
To demonstrate this matter, we pruned the widely utilized open-source model ResNet50 Gender Classifier~\cite{RasbtDeeplearningmodelsCollection} using PyTorch's official LTH pruning algorithm package~\cite{TorchNnPyTorch}. 
Subsequently, we assessed the fairness and accuracy of the original model on the CelebA dataset~\cite{liu2015faceattributes} concerning its sparser versions obtained through LTH-based pruning, with pruning ratios set at 0.9, 0.95, and 0.99, respectively.
The accuracy, CWV, and MCD are recorded for each model, and the results are presented in \autoref{fig:lthfair}.
The bars in the figure are pattern-coded, with diagonal-hatching representing the original model, and crosshatching, dotted pattern, and plus sign pattern corresponding to pruning rates of 0.9, 0.95, and 0.99~(i.e. sparsity of 0.10, 0.05 and 0.01), respectively. 
As depicted in the figure, while not leading to a significant decrease in model accuracy, an increase in model sparsity results in a significant reduction in fairness performance.

This observation highlights that the LTH approach, being primarily focused on identifying a winning ticket that can retain its capacity toward test accuracy throughout training, tends to overlook potential fairness issues.
Consequently, a pertinent question arises: \textit{Is it feasible to identify a winning ticket that excels in both accuracy and fairness concurrently?}

%% file: misc/LTHfair.tex
\begin{figure*}
    \centering     
    \scalebox{1.0}{
  \begin{subfigure}[figure1]{0.33\linewidth}  
        \centering
    \includegraphics[height=3.5cm]{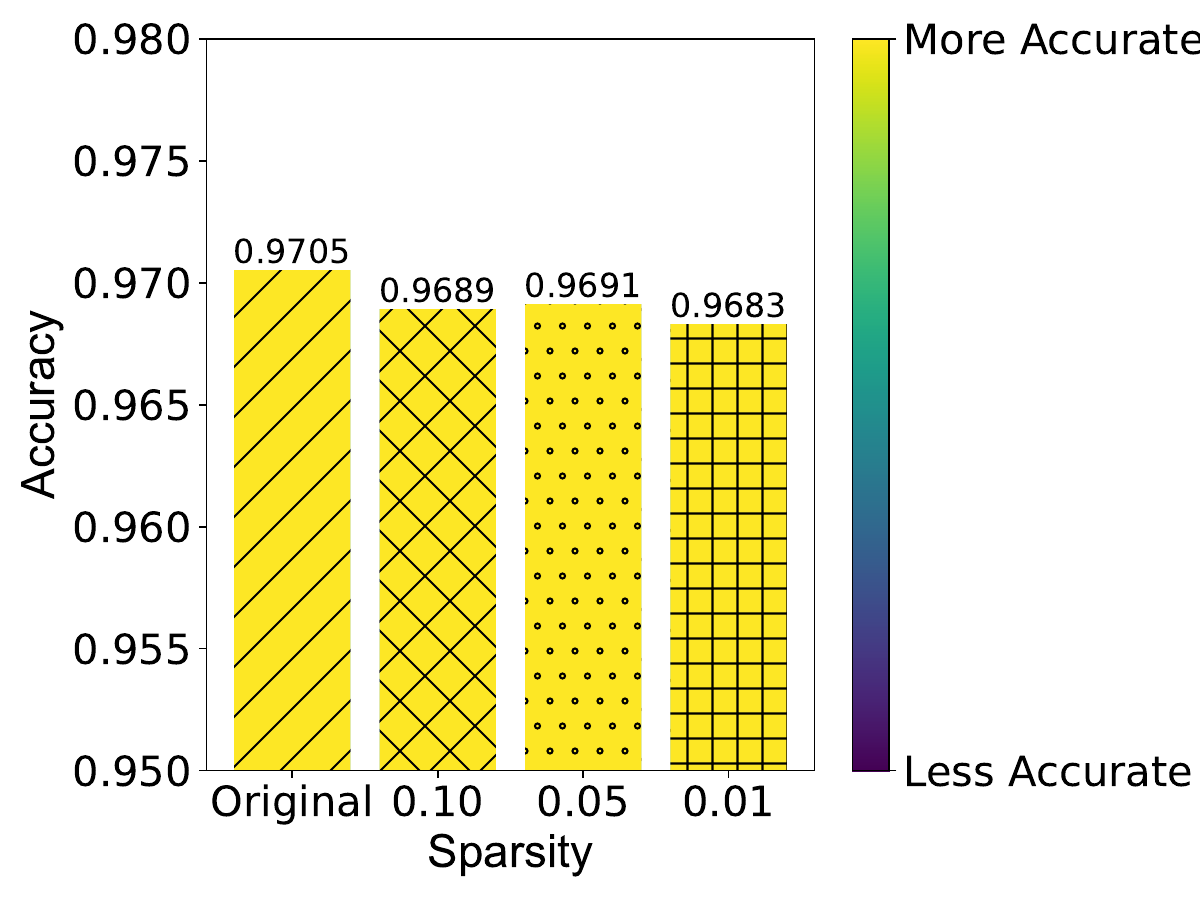} 
      \caption{Accuracy}
      \label{fig:moti_acc}
  \end{subfigure}
  \begin{subfigure}[figure2]{0.33\linewidth}
        \centering
    \includegraphics[height=3.5cm]{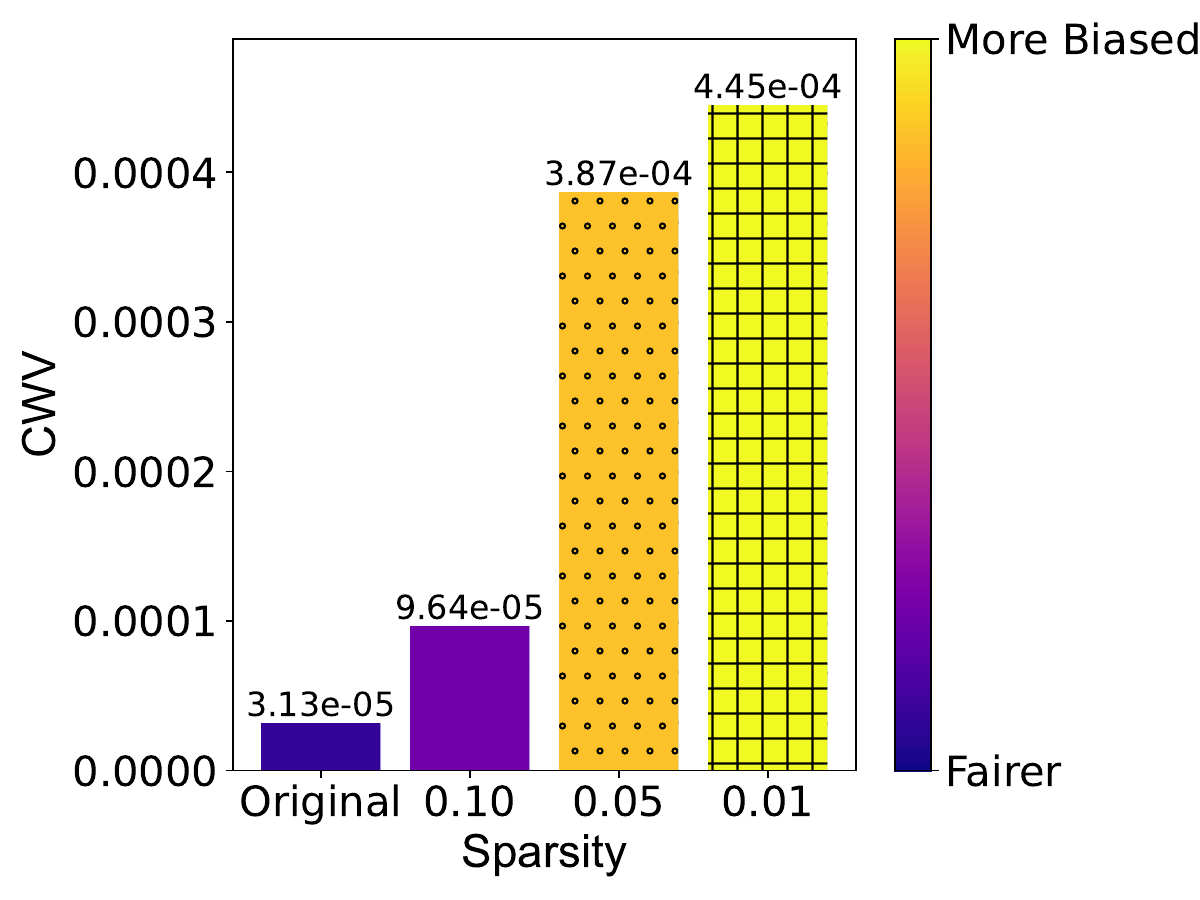} 
      \caption{CWV~(larger value, higher degree of bias)}
      \label{fig:moti_cwv}
  \end{subfigure}
  \begin{subfigure}[figure3]{0.33\linewidth}
    \centering
  \includegraphics[height=3.5cm]{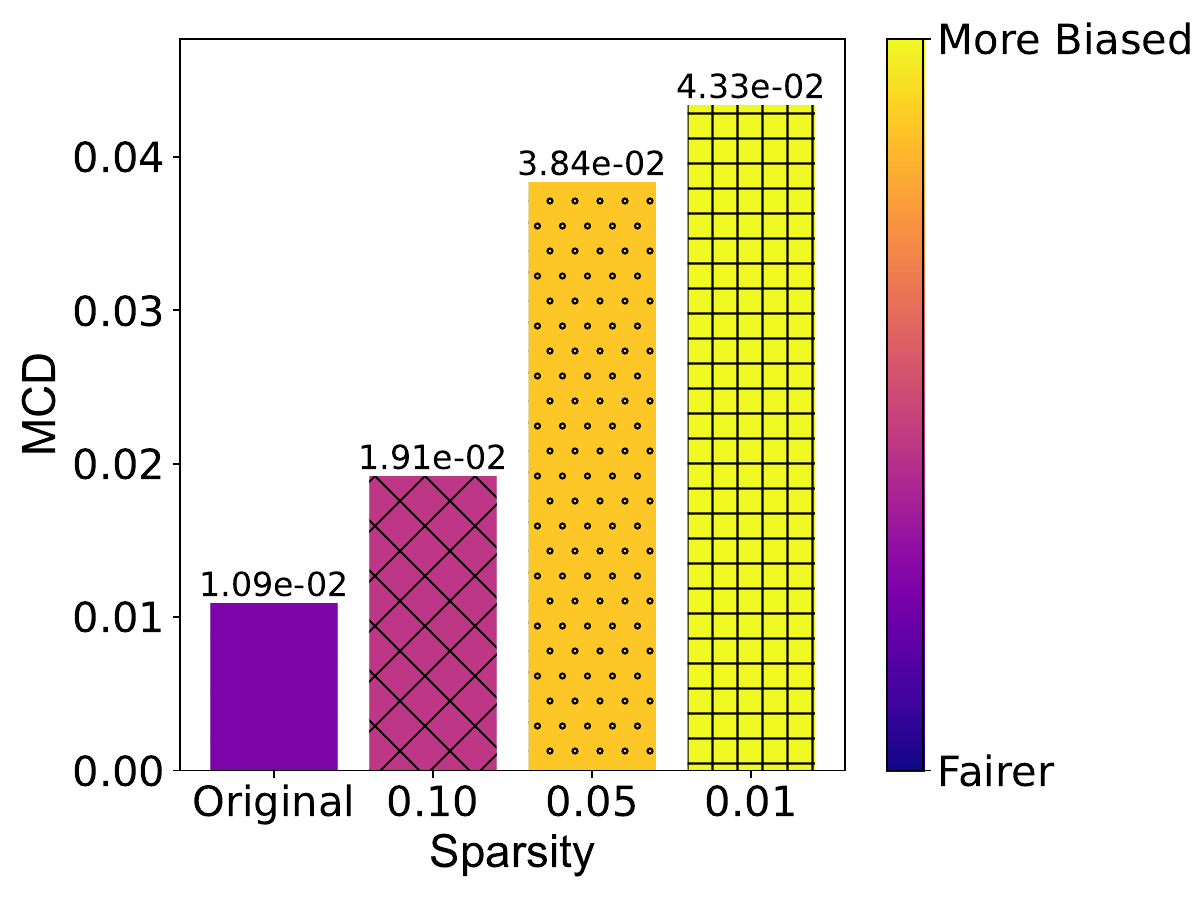} 
    \caption{MCD~(larger value, higher degree of bias)}
    \label{fig:moti_mcd}
  \end{subfigure}
  }
  \caption{Performance of ResNet50 Gender Classifier and its sparse versions after LTH-based pruning.}
  \label{fig:lthfair}
  \vspace{-14pt}
\end{figure*}

%% file: body/design.tex
\section{System Design}\label{sec:design}

\subsection{Problem Statement} \label{sec:ps}

In this paper, we aim to develop an automated model pruning framework that concurrently addresses both pruning rate and fairness considerations.
For a given initialized complex network \(f(\theta_{0})\) and training epochs \(E\), the \(f(\theta_{0})\) can be normally trained into a model \(f(\theta)\) with \(E\) epochs.
We focus on how to select a ticket to satisfy the specified sparsity metric \(\Omega\), and retrain this ticket into a final sparse model \(f(\theta^{*})\) by a training method \(T\).
The \(f(\theta^{*})\) should attain a high level of fairness, while fulfilling the availability requirement stated by the sparsity metric \(\Omega\) and accuracy specifications \(\epsilon\) as \autoref{eq:prunerate} and \autoref{eq:pruneperformance}.
Formally, our objective is:
\begin{align}\label{eq:targetproblem}
    \min_{m, T} \left[ d(f(\theta^*)) - d(f(\theta)) \right]\quad\quad\quad \quad\\
    \quad \text{s.t.} \quad \frac{|\theta^*|}{|\theta|} \leq \Omega
    \quad \text{Acc}(f(\theta)) - \text{Acc}(f(\theta^{*})) \leq \epsilon \nonumber
\end{align}
where the \(d(\cdot) \in \{CWV, MCD \}\) is a measure of bias degree referred to \autoref{sec:bgfairness}, which represents a higher fairness with lower indicator value.
The \(|\theta|\) and \(|\theta^*|\) calculate the number of parameters, while \(Acc(\cdot)\) calculates the accuracy of the model on the test dataset.
Note the pruning rate is \((1-\Omega)\).
Our objective is to enhance the fairness of the pruning model as much as possible while preserving its performance (i.e., accuracy). 
This involves finding a balance between accuracy and fairness, making trade-offs when necessary.

\subsection{Root Cause Analysis} \label{sec:rca}

In pursuit of an accurate and fair ticket, it is essential to consider the trade-off involved in optimizing both aspects.
We believe that trade-off arises at two primary levels: first, in determining what components of the model should be removed during pruning~(referred to as \textit{ticket selection}); and second, in devising an appropriate training strategy for the model after the removal~(referred to as \textit{ticket training}).
Below, we conduct an exhaustive analysis of the factors contributing to fairness issues in both ticket selection and ticket training.

\subsubsection{Ticket Selection}\label{sec:ticketselection}

Fairness and accuracy may conflict in the search for tickets. 
For DNN models, the ticket selection and pruning process involves removing redundant neurons.
Typically, high-dimensional features extracted from samples are stored in neurons as combinations. 
Thus, ticket selection is equivalent to feature selection on the dataset, which represents a form of high-dimensional feature selection for a sample.   
However, this selection process may introduce bias.
As previously discussed, LTH methods predominantly prioritize accuracy, potentially leading to the identification of tickets that lack fairness. 
Such tickets may resort to taking shortcuts instead of genuinely learning the intended solution.
For instance, in a face recognition software deployed in a predominantly white neighborhood, a ticket may only use skin color as the determinant of whether a visitor is a local resident, resulting in reduced generalization and fairness~\cite{geirhosShortcutLearningDeep2020a}.
Conversely, exclusively focusing on fairness during ticket identification may result in a decline in accuracy. 
Intuitively, if a face recognition software blocks out all features related to skin color to be fair, its face recognition performance is likely to suffer considerably.
Hence, a trade-off must be carefully considered throughout the process of ticket selection to strike a balance between fairness and accuracy.

\input{misc/maskxai.tex}

We study the association of neurons to features by employing an interpretable AI technique known as GradCAM~\cite{selvaraju2017grad}.
The GradCAM heatmap visually represents the regions in an image that the deep learning model deemed most influential in making its prediction, wherein ``hot'' regions indicate areas of significant influence on predicting the object's class membership.
The primary objective of using the GradCAM heatmap is to visualize the regions of the input image that the model's attention is directed towards under varying masks, shedding light on the model's feature focus during its decision-making process.
The experiment is based on the ResNet-50 model, which was trained on the PubFig dataset~\cite{kumar2009attribute}.
\autoref{fig:maskxai} shows how the neuron mask affects the model's selection of features.
In the figure, regions highlighted in orange depict areas where the model exhibits high attention, while regions colored in blue represent areas with lower model attention.
Note that (a) represents the original image, while (b), (c), and (d) correspond to the saliency maps of the original model, the model after random mask application, and the model after \sys pruning, respectively.
Obviously, the region of interest of the model undergoes a shift as the neurons involved in decision-making are different.
The region of interest of the model after random mask depicted in (c) exhibits a significant shift compared to the original model (b), diverting its focus from facial features to other attributes that are unrelated to face classification but may carry discriminatory information~(e.g., hair color, clothing, etc.). 
In contrast, the model after \sys pruning appears to maintain a closer resemblance to the original model, which likely indicates that the model prioritizes the region containing features essential for face classification. 
This observation underscores the substantial influence that the judicious selection of neurons can have on the fairness of the model.
We further corroborate this through subsequent empirical experiments~(see~\autoref{sec:rq3}).


\subsubsection{Ticket Training}\label{sec:tickettraining}


Even upon obtaining a ticket that strikes a balance between accuracy and fairness, there is no guarantee of successfully ``claiming a prize'', that is, training a model that attains the upper limit of its theoretical performance.
There are two potential issues related to the training process. 
First, if the model excessively focuses on learning the features of existing training samples, it may treat some of these specific features as if they are universally representative of all potential samples, leading to overfitting. 
In such cases, the model becomes too tailored to the training data and may not generalize well to new, unseen samples, thus introducing bias.
Conversely, if the model's learning capacity is too limited, it may not adequately capture the general features of the samples. 
This situation, known as underfitting, results in a failure to learn critical patterns from the data, leading to poor performance on both training and unseen samples.
Achieving an optimal balance between overfitting and underfitting is essential for obtaining a model that generalizes well and demonstrates accurate predictions on new and diverse samples.

\input{misc/fitting.tex}

We conduct a series of training experiments on the CIFAR-100 dataset, utilizing the same ticket~(model prototype) acquired through standard LTH pruning with a pruning rate of 0.95. 
The experiments involve training the model for 50 and 150 epochs~(underfitting model), 250 epochs~(normally fitting model), and 500, 750, and 1000 epochs~(overfitting model). 
Throughout all experiments, a uniform learning rate of 0.03 is employed.
Then we evaluate the accuracy and fairness of these trained models.
\autoref{fig:fitting} compares the accuracy, CWV, and MCD of these models.
As we can see, the normal model exhibits the most favorable performance in both fairness and accuracy, whereas the underfitting model demonstrates the poorest performance. 
This highlights that, despite employing the same ticket, the training process significantly influences the final performance of the retrained model.

\subsection{System Overview} \label{sec:overview}

Through our analysis, it becomes evident that existing methods encounter challenges in finding tickets that simultaneously achieve both accuracy and fairness while successfully training a model.
In response, our proposed framework, \sys, introduces novel \textit{conflict-detection-based mask generation} and \textit{training refinement} techniques to obtain accurate and fair sparse models. We first locate the root cause, namely neurons exhibiting conflicts between fairness and accuracy during optimization, and then generating masks that block out these neurons to get a sparse model and train it by a comprehensive strategy to improve its fairness performance. 
The primary objective of mask generation is to ensure that the model focuses on the most relevant features during the learning process.
The training refinement is pivotal in ensuring that the model not only captures essential features but also achieves equitable outcomes across different groups or classes.
\input{misc/overview.tex}

\autoref{fig:overview} presents the workflow of \sys.
Following the conventional LTH approach, we commence by training randomly initialized models while recording the gradient values of each neuron during each round of model training with respect to accuracy loss and fairness loss. 
Subsequently, we employ these gradient values to conduct an analysis and identify neurons that present conflicts between fairness and accuracy optimization objectives. 
Next, we generate masks corresponding to these identified neurons and proceed to prune them to address the conflicts. 
Finally, we undertake a training refinement step on the sparse model to achieve an optimized model.
We dynamically adjust the learning rate during training to enhance the model's acquisition of more nuanced features, and if the trained model's fairness lags behind the original model, we iteratively retrain it with weights reloaded from earlier rounds until further improvement is unattainable.

The overall algorithm of \sys is presented in \autoref{alg:overview}, denoted as procedure \textit{FixModel}.
It takes a base model \(f(\theta_0)\), the pre-determined sparsity metric \(\Omega\) and the number of training epochs \(E\) as inputs, and outputs an optimized model \(f(\theta^{*})\).
Firstly,  \sys engages in a standard training process on the base model \(f(\theta_0)\), recording the accuracy loss \(L_a\) and the fairness loss \(L_f\) for each epoch~(line 2-7).
This meta-data collection aims to acquire insights into the optimization directions concerning accuracy and fairness during training, facilitating subsequent conflict analysis.
Next, \sys utilizes this information to identify the mask corresponding to a ticket that satisfies both accuracy and fairness~(line 8).
Traditional LTH methods often prioritize accuracy, leading to subpar fairness performance.
In contrast, \sys guides the model to optimize towards the direction of conjugate optimization by analyzing the directions of fairness and accuracy optimization, thereby enhancing the likelihood of discovering a model that is both accurate and fair.
Finally,  \sys employs training refinement (lines 9-11) to optimize the performance of the discovered tickets.
Through this refined training, \sys guarantees the preservation of model accuracy while concurrently striving to uphold fairness.

\input{misc/algorithm.tex}

\subsection{Mask Generation} \label{sec:mask}


To address this challenge mentioned in \autoref{sec:ticketselection}, we propose a novel mask generation approach termed \textit{conflict-detection-based mask generation}.
Specifically, we collect the gradients corresponding to the fairness and accuracy of each neuron during training the model in the normal way, and then locate neurons with optimization conflicts and generating masks to remove them.

Prior to delving into our approach, let us first introduce the precise definitions of the accuracy loss function~(denoted as \(l_a\)) and the fairness loss function~(denoted as \(l_f\)) employed in our methodology.
The \(l_a\) and \(l_f\) can be formulated as:
\begin{equation}\label{eq:accloss}
    l_a=-\sum_{c=1}^C \log \frac{\exp(x_{n,c})}{\sum_{i=1}^C\exp(x_{n,i})} y_{n,c}
\end{equation}
\begin{equation}\label{eq:fairloss}
    l_f=-\sum_{c=1}^C w_c\log \frac{\exp(x_{n,c})}{\sum_{i=1}^C\exp(x_{n,i})} y_{n,c}
\end{equation}
In the context of the provided equations, \(C\) denotes the number of classes, \(x\) represents the input data, \(y\) denotes the target and \(w\) corresponds to the weight.
Here, we set the weights as \(\frac{1}{\hat{acc}_c}\), where \(\hat{acc}_c\) denotes the accuracy of the previous iteration of the model.
For each respective class, we take this value as an estimate of the current iteration's accuracy.  
The choice of weights follows an inverse proportionality to the accuracy achieved in the prior iteration for individual classes.



Initially, during the standard training process, we collect the gradient values of each neuron pertaining to both the fairness and accuracy loss functions for each training round~(line 4-7). 
This information serves as the foundation for evaluating potential optimization conflicts. 
Then we compute the sum of these two values for every round, weighted by a certain percentage \(\gamma\) to ensure that both values are on the same order of magnitude~(line 12-18).
Here, the value of \(\gamma\) is predetermined based on the outcomes of hyperparameter experiments, as discussed in \autoref{sec:rq4}.
The resulting sums are sorted to derive a reference value, denoted as  conflict degree \(l\), delineate the degree to which the optimization objectives of the two loss functions are in conflict during each respective round~(line 19).
Next, we find out the neurons with the highest degree of conflict and increment their respective count~(line 20-21). 
By summing the count values of each neuron across all rounds, we identify the neurons with the highest number of conflict rounds~(line 22-24). 
These neurons are then removed based on the sparsity metric to obtain the desired mask for our approach.

Through an examination of the metadata from the training process, we identify neurons whose optimization exhibited conflicts between the fairness and accuracy objectives. 
This analysis aid in the discovery of potential models that are more inclined to achieve fairness-accuracy conjugate optimization by eliminating these neurons.
However, merely identifying such a sparse model prototype~(ticket) is insufficient; it remains crucial to consider how to effectively train this prototype into a fully functional and high-performing model~(claim a prize).
The optimization and validation of ticket training are imperative.

\subsection{Training Refinement} \label{sec:training}

Upon identifying a ticket, our task is only halfway completed. 
While this ticket holds the potential to serve as a fair and accurate model, careful training is indispensable to enable it.
Traditional LTH methods involve loading the ticket with the initial base model weights \(\theta_0\) and subsequently performing retraining, akin to the initial training.
However, such an approach lacks optimization steps and neglects performance validation of the trained model. 
Thus, there is a need for a more comprehensive training strategy that includes optimization techniques and robust performance validation to ensure the model attains its full potential.

To address this, we undertake a comparison of various training policies and optimize the training procedure employing a technique termed \textit{training refinement}.
In detail, we commence the training refinement process by reloading the initial weights into the pruned model. 
Subsequently, we conduct model training using a decreasing learning rate strategy to facilitate learning as many essential features as possible while avoiding overfitting. 
The objective is to strike a balance between adequately capturing necessary patterns in the data and ensuring the model's ability to generalize effectively to new, unseen samples.
Ultimately, we validate the retrained model. 
If the model's fairness performance falls below that of the original model, we initiate a feedback loop by returning to the first step. 
During this process, instead of reloading the base model initialization weights, we reload the weights after completing a specified number~(typically a small single or double digit number predetermined by empirical considerations, see \autoref{sec:rq3})
of training epochs and recommence the training procedure. 
This is because sparse subnetworks that remain stable to the optimization noise in the early stages of training are more likely to be winning tickets~\cite{frankle2020linear}.
We iterate through the aforementioned steps until fairness or accuracy decreases above the threshold value \(\epsilon\) or ceases to improve further.
It increases the likelihood of identifying and training winning tickets to the desired levels of accuracy and fairness.
By following the aforementioned process, we mitigate potential issues such as convergence failure or convergence to a local optimum, which may arise during the training process. 
Our approach endeavors to ensure that the previously obtained accurate and fair tickets can be effectively employed to train a high-performing model.

%% file: misc/maskxai.tex
\begin{figure}[t]
    \centering
    \footnotesize
    \scalebox{0.9}{
    \begin{subfigure}[t]{0.22\linewidth}
            \centering     
            \footnotesize
            \includegraphics[width=\textwidth]{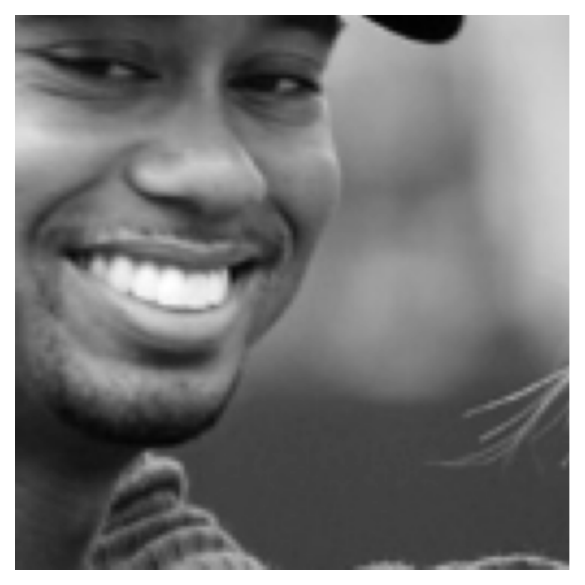}
        \end{subfigure}
        \begin{subfigure}[t]{0.22\linewidth}
            \centering
                   \footnotesize
                   \includegraphics[width=\textwidth]{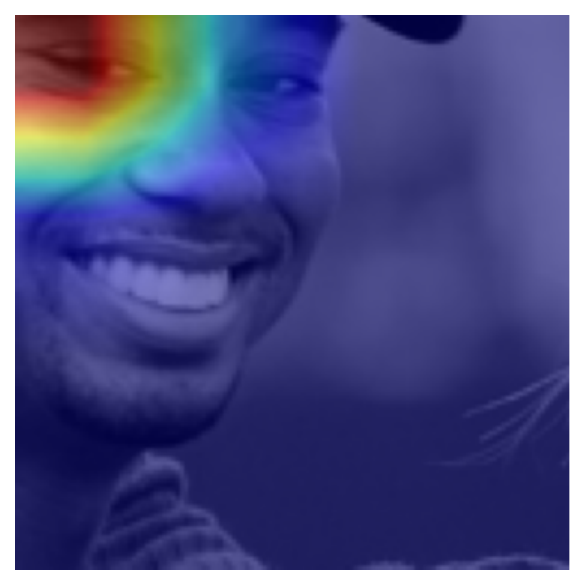}
        \end{subfigure}
        \begin{subfigure}[t]{0.22\linewidth}
            \centering     
            \footnotesize
            \includegraphics[width=\textwidth]{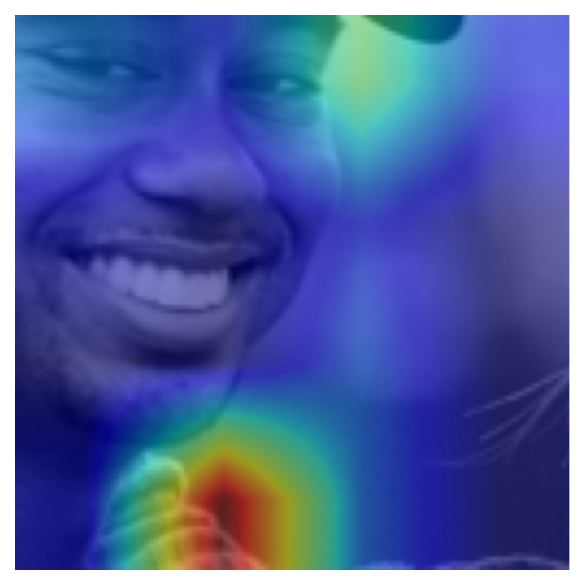}
        \end{subfigure}
        \begin{subfigure}[t]{0.22\linewidth}
            \centering     
            \footnotesize
            \includegraphics[width=\textwidth]{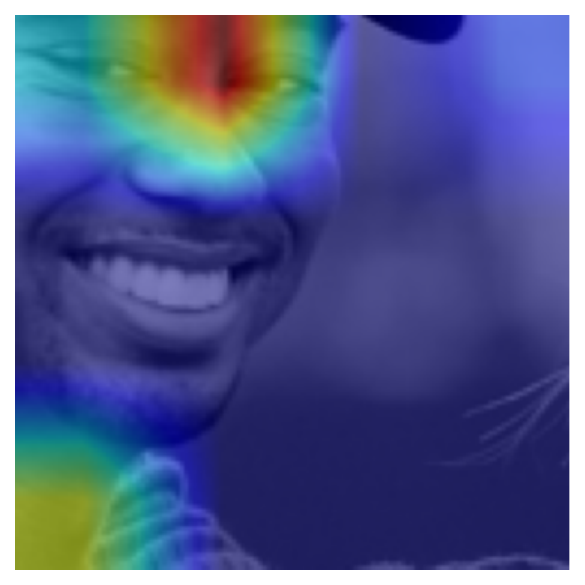}
        \end{subfigure}
        }

        \scalebox{0.9}{
        \begin{subfigure}[t]{0.22\linewidth}
            \centering
                   \footnotesize
                   \includegraphics[width=\textwidth]{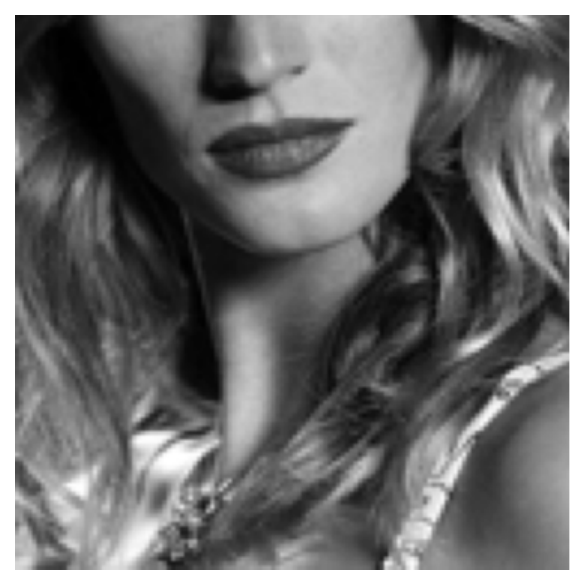}
                   \caption{}\label{f:xai_input}
        \end{subfigure}
        \begin{subfigure}[t]{0.22\linewidth}
            \centering     
            \footnotesize
            \includegraphics[width=\textwidth]{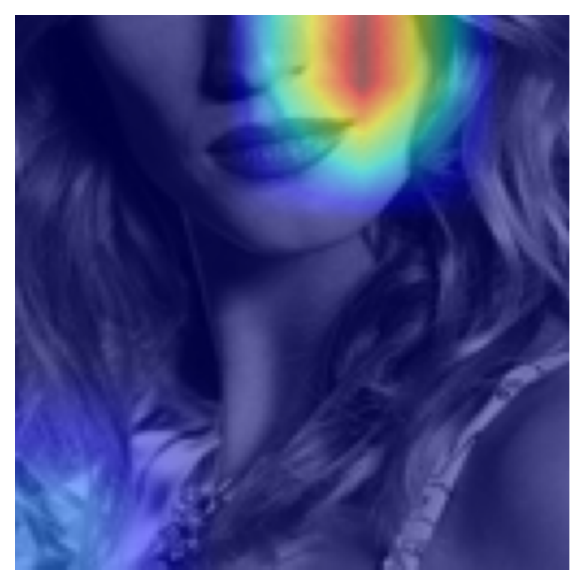}
                   \caption{}\label{f:xai_orim}
        \end{subfigure}
        \begin{subfigure}[t]{0.22\linewidth}
            \centering
                   \footnotesize
                   \includegraphics[width=\textwidth]{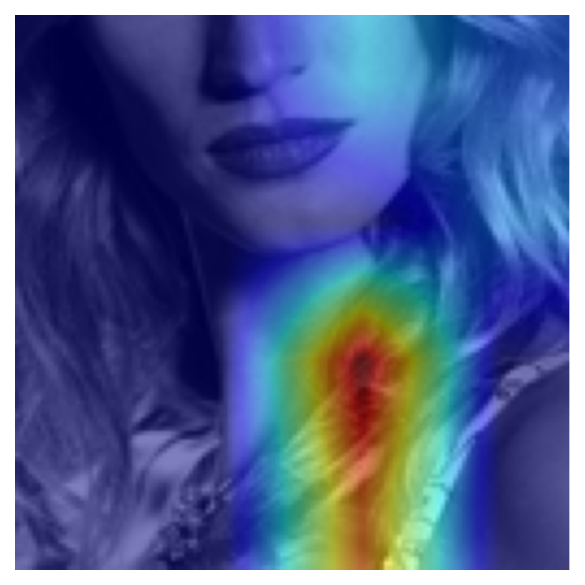}
                   \caption{}\label{f:xai_random}
        \end{subfigure}
        \begin{subfigure}[t]{0.22\linewidth}
            \centering     
            \footnotesize
            \includegraphics[width=\textwidth]{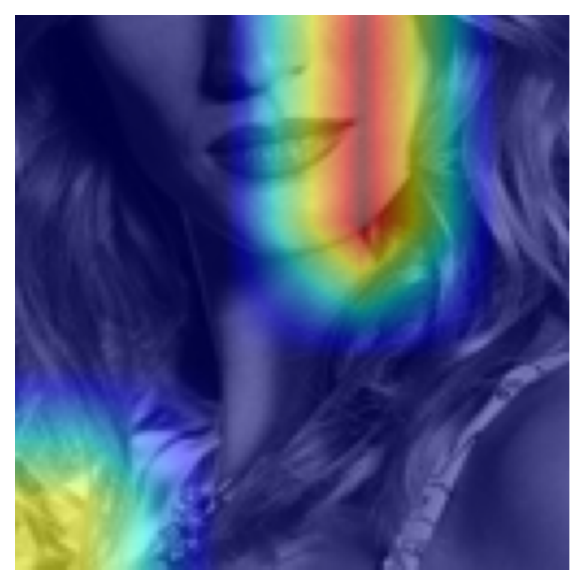}
                   \caption{}\label{f:xai_lth}
        \end{subfigure}
        }
    \caption{GradCAM heatmap for different models. (a) is the original image; (b)-(d) are the results of the original model, the model after random mask application and the model after \sys pruning.}\label{fig:maskxai}
    \vspace{-15pt}
\end{figure}

%% file: misc/fitting.tex

\begin{figure}
    \centering   
    
  \begin{subfigure}[figure1]{0.49\linewidth}  
        \centering
    \scalebox{1.0}{
    \includegraphics[height=3cm]{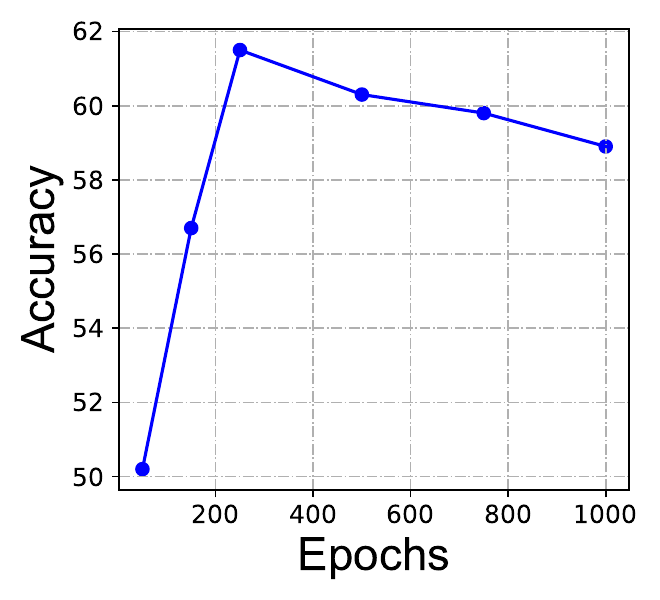} 
    }
      \caption{Epochs-accuracy}
      \label{fig:fitting_1}
  \end{subfigure}
  \begin{subfigure}[figure2]{0.49\linewidth}
        \centering
    \scalebox{1.0}{
    \includegraphics[height=3cm]{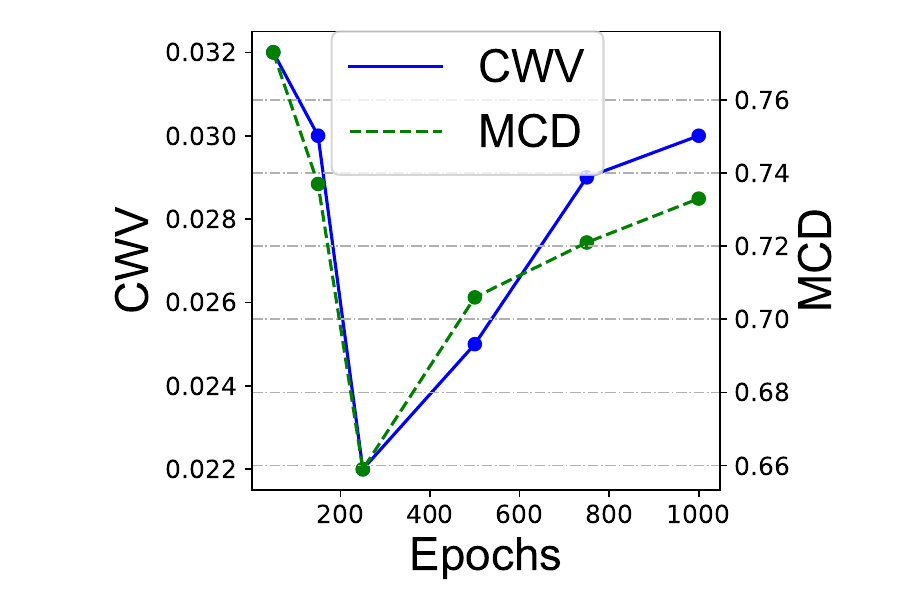}
    }
      \caption{Epochs-CWV\&MCD}
      \label{fig:fitting_2}
  \end{subfigure}
  
  \caption{Average model performance on CIFAR-100 dataset for different training epochs setup.}
  \label{fig:fitting}
  \vspace{-20pt}
\end{figure}

%% file: misc/overview.tex
\begin{figure}[h]
    \centering
    \scalebox{0.95}{
    \includegraphics[trim={1cm 8.5cm 10.2cm 12.5cm},clip,width=\linewidth]{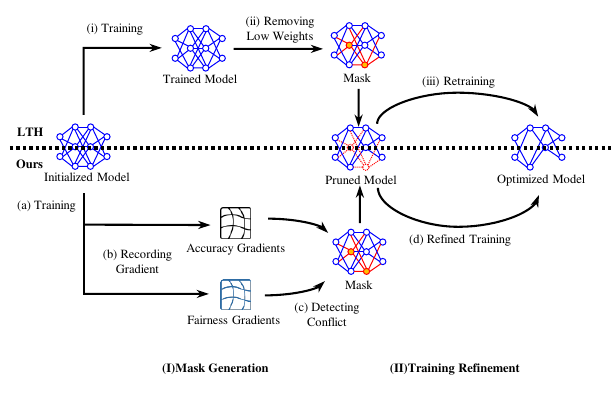}
    }
    \caption{Overview of {\sys}.}
    \label{fig:overview}
    \vspace{-10pt}
\end{figure}

%% file: misc/algorithm.tex
\begin{algorithm}[t]
    \scriptsize
    \caption{\sys Algorithm}
    \label{alg:overview}
    \begin{algorithmic}[1]
        \Require \(f(\theta_0)\): base model with random initialized weight
        \Require \(\Omega\): sparsity metric
        \Require \(E\): number of training epochs
        \Ensure \(f(\theta^{*})\): optimized model
        \Procedure{FixModel}{}
        \State $L_a \gets []$ \Comment{\textit{list of model accuracy loss}}
        \State $L_f \gets []$ \Comment{\textit{list of model fairness loss}}
        \For{$i \gets 0 \  to \ E $}
        \State $f(\theta_{i+1}) \gets Training(f(\theta_i))$
        \State $L_a[i] \gets AccLoss(f(\theta_{i+1}))$
        \State $L_f[i] \gets FairLoss(f(\theta_{i+1}))$
        \EndFor
        \State $m \gets GenerateMask(L_a,L_f)$
        \For{$i \gets 0 \  to \ E $}
        \State $f(\theta_{i+1} \odot m) \gets RefinedTraining(f(\theta_i \odot m))$
        \EndFor
        \State $f(\theta^{*}) \gets f(\theta_E \odot m)$
        
        \Return $f(\theta^{*})$
        \EndProcedure   

        \item[]
        \Require \(L_a\): list of accuracy loss for each epoch
        \Require \(L_f\): list of fairness loss for each epoch
        \Require \(\gamma\): hyperparameter used to control the loss weight
        \Require \(\eta\): hyperparameter used to control the neuron sorting ratio
        \Ensure \(m\): mask corresponding to fair and accurate ticket
        \Procedure{GenerateMask}{}
        \State $L_c \gets []$ \Comment{\textit{list of model weighted loss}}
        \State $count \gets []$ \Comment{\textit{Record the number of conflicts for each neuron}}
        \For{$i \gets 0 \  to \ |\theta| $}
        \State $count[i] \gets 0$
        \EndFor
        \For{$i \gets 0 \  to \ E $}
        \State $L_c[i] \gets L_a[i]+\gamma \times L_f[i]$
        \State $l \gets Sort(L_c[i].neuronvalue)$
        \For{$j \gets 0 \  to \ \eta \times |\theta| $}
        \State $count[l[j].index] \gets count[l[j].index]+1 $
        \EndFor
        \EndFor
        \State $countsorted \gets Sort(count)$
        \For{$i \gets 0 \  to \ \Omega \times |\theta| $}
        \State $Append(m,countsorted[i].neuron)$
        \EndFor
        \Return $m$
        \EndProcedure  
    \end{algorithmic}

\end{algorithm}

%% file: body/Evaluation.tex
\section{Evaluation}\label{sec:eval}

We aim to answer the following research questions through our experiments:

\noindent \textbf{RQ1:} 
Can \sys effectively claim the winning ticket?

\noindent \textbf{RQ2:} 
How efficient does \sys find and claim the winning ticket?

\noindent \textbf{RQ3:} 
How do mask generation and training policies affect the performance of \sys?

\noindent \textbf{RQ4:} 
What is the impact of different configurable parameters in \sys?

\vspace{-8pt}

\subsection{Setup}\label{sec:setup}

\subsubsection{Software and Hardware}
The prototype of \sys is implemented on top of PyTorch 2.0.
We conduct our experiments on a server with 64 cores Intel Xeon 2.90GHz CPU, 256 GB RAM, and 4 NVIDIA 3090 GPUs running the Ubuntu 16.04 operating system.

\vspace{-5pt}

\subsubsection{Datasets}
We evaluate \sys on five popular datasets: CIFAR-100~\cite{CIFAR100Datasets}, TinyImageNet~\cite{le2015tiny}, CelebA~\cite{liu2015faceattributes}, LFW~\cite{LFWTech}, and Moji~\cite{blodgett2016demographic}.

\begin{itemize}
    \item {\bf CIFAR-100} is a widely-used colored image dataset used for object recognition. It contains 60,000 32x32 color images in 100 classes, of which 50,000 are training images and the rest are test images.
    \item {\bf TinyImageNet} is a subset of the ILSVRC2012 classification dataset.
    It consists of 100k colored images of 200 classes, and all images have been down-sampled to \(64\ast64\ast3\) pixels.
    \item {\bf CelebA} is a large-scale face attributes dataset with more than 200K celebrity images.
    The images in this dataset cover large pose variations and background clutter.
    We perform face gender classification task on this dataset.
    \item {\bf LFW} is a collection of over 13,000 face photographs sourced from the web.
    Many groups are underrepresented in this dataset, such as children, babies, individuals over the age of 80, and women.
    Therefore, LFW is suitable for evaluating model performance in the data imbalance setting.
    We perform face gender classification task on this dataset.
    \item {\bf Moji} contains more than 100,000 tweets written in either ``Standard American English'' or ``African American English'', annotated with positive or negative sentiment.
    We perform text sentiment classification task on this dataset.
\end{itemize}

\vspace{-5pt}

\subsubsection{Models}

For a fair comparison with baseline methods, we reproduce the baselines and use the same models and setup on each method in the experiments.
We train the 50-layer ResNet model (i.e., ResNet50)~\cite{heDeepResidualLearning2016} and the VGG16~\cite{simonyan2014very} with SGD and set the batch size to 32.
In the training process, the learning rate starts from 0.1 and reduces to 1/10 of the previous learning rate after 100, 150 and 200 epochs (250 in total).
When pruning models, we set the sparsity to 0.05, i.e., remove 95\% of the parameters in the model.
Random cropping, horizontal flip, and normalization are adopted for data augmentation.

To assess the performance of the \sys in natural language processing tasks, we conducted experiments on the BERT model~\cite{devlin2018bert}.
We finetune the BERT model with AdamW, set the batch size to 32 and the learning rate to 5e-5.

\vspace{-5pt}

\subsubsection{Baselines} 

We compare \sys with other state-of-the-art model pruning methods, including Magnitude Pruning~\cite{han2015learning}, Standard LTH~\cite{frankle2018lottery}, SafeCompress~\cite{zhu2022safety}, and FairScratch~\cite{tangFairScratchTickets2023}.
In experiments, We run the baseline methods with the default settings which are recommended in their papers and open-source code and record their performance.

\noindent
{\bf Magnitude Pruning.} Han et al.~\cite{han2015learning} trained a model to learn which neuronal connections are important, and then they pruned those unimportant ones and fine-tuned the target model.
Their method can effectively reduce the number of parameters of the target model by an order of magnitude without affecting their accuracy.

\noindent
{\bf Standard LTH.} Frankle et al.~\cite{frankle2018lottery} proposed the Lottery Ticket Hypothesis (i.e., Standard LTH in this paper) that a DNN contains subnetwork (e.g., the winning ticket) that is able to match the test accuracy of the original network after training.
They found the winning ticket by masking the original model, thereby reducing the number of parameters of the model while ensuring accuracy.


\noindent{\bf SafeCompress.} Zhu et al.~\cite{zhu2022safety} proposed SafeCompress, which is a test-driven sparse training framework for safe model compression.
SafeCompress first prunes the big model to an initial sparse model, and then iteratively compresses the sparse model with various compression strategies.
We reuse their open-source code by replacing the safety metric with fairness metrics.

\noindent{\bf FairScratch.} Tang et al.~\cite{tangFairScratchTickets2023} proposed a fairness learning framework for computer vision system without weight training.
It tries to find a suitable mask in the initialized original model, whose corresponding subnetwork has a high performance without training.

First three works designed model pruning and compression methods from the perspective of model performance and security, which effectively enhance the accuracy and safety of pruned models.
However, they all raise fairness issues.
The reason is that none of them consider the impact of pruning on model bias and fairness, nor use any metrics as guidance. 
In fact, during the pruning process, they may prune some neurons that are strongly associated with fairness in order to ensure the accuracy of the model, thereby introducing bias into the model, which makes it difficult for the model to achieve high fairness.
FairScratch, on the other hand, does not consider the increased difficulty of identifying winning tickets without weight training, particularly in scenarios with high compression rates, which limits its performance.


\vspace{-5pt}

\subsubsection{Evaluation metrics}
We compare \sys with the baselines on five metrics: accuracy, precision, recall, CWV and MCD~(see~\autoref{sec:bgfairness}).

\input{body/rq1.tex}
\input{body/rq2.tex}

\input{body/rq3.tex}
\input{body/rq4.tex}

%% file: body/rq1.tex
\subsection{RQ1: Effectiveness in Finding and Claiming Winning Ticket}
\label{sec:rq1}

\input{tftex/rq1_tab}


\noindent
{\bf Experiment Design}:
To evaluate the effectiveness of \sys in finding and claiming in the winning ticket to fix the bias problems, we use the following three baseline pruning methods and \sys to conduct experiments on ResNet50, VGG16 and BERT on the five datasets, which are mentioned in~\autoref{sec:setup}.
In the experiment, we fixed the training settings and dataset of the model and ran the training 10 times to ensure the reliability of the results and reduce the randomness.
We compare the performance between \sys and baseline methods in terms of both utility and fairness.

\noindent
{\bf Results}:
The comparison results are presented in~\autoref{tab:rq1-1}.
The first column shows two datasets and the second column lists four methods in experiments including \sys.
We record the model performance of the original models that have not been pruned, which is represented as `Origin' in the second column.
The remaining columns list the performance of the pruned model, including accuracy (Acc.), precision (Precis.), recall, class-wise variance (CWV), and maximum class-wise discrepancy (MCD).
The best results among different pruning methods are highlighted in bold.

The experiment results demonstrate that the winning tickets exist and \sys can effectively find and claim these lottery tickets.
On the one hand, \sys can effectively alleviate the fairness bias in model pruning and the ticket \sys found has better performance in fairness than other baselines.
The last two columns of~\autoref{tab:rq1-1} show the fairness improvement of \sys on these datasets.
The models pruned by \sys achieve the lowest CWV~(indicates the highest fairness) among all models, which exceeds the Magnitude Pruning by 32.52\%, 53.71\%, and 71.92\%; Standard LTH by 35.60\%, 56.78\%, and 71.91\%; SafeCompress by 60.87\%, 58.33\%, and 61.49\%; and FairScratch by 43.02\%, 60.11\%, and 56.57\% on pruning ResNet50, VGG16, and BERT respectively.
On average, \sys improves CWV by 49.05\%, 54.76\%, 60.23\%, and 53.23\% compared to state-of-the-art baselines, namely Magnitude Pruning, Standard LTH, SafeCompress, and FairScratch.
In terms of MCD, the \sys also has advantages in alleviating the class-wise discrepancy over other baselines.
MCD of \sys exceeds the Magnitude Pruning by 43.81\%, 44.79\%, and 66.72\%; Standard LTH by 48.25\%, 46.88\%, and 66.67\%; SafeCompress by 57.81\%, 52.78\%, and 52.21\%; and FairScratch by 46.20\%, 49.18\%, and 49.17\% on pruning ResNet50, VGG16, and BERT respectively.
On average, \sys improves MCD by 51.77\%, 53.93\%, 54.27\%, and 48.18\% compared to state-of-the-art baselines, namely Magnitude Pruning, Standard LTH, SafeCompress, and FairScratch.
Although the MCD result of the model pruned by \sys is slightly behind the original model, it has already significantly exceeded the four baselines.

On the other hand, \sys effectively preserves and even improves the model accuracy, precision, and recall when pruning models and achieves the highest utility among all methods.
From~\autoref{tab:rq1-1}, we can observe that \sys outperforms the state-of-the-art methods on accuracy, precision and recall when pruning models.
Compared to the non-pruned models (i.e., original model), Magnitude Pruning degrades 8.51\%, 8.75\%, and 4.41\%; Standard LTH degrades 3.03\%, 3.25\%, and 1.75\%; SafeCompress degrades 8.79\%, 9.05\%, and 2.76\%; and FairScratch degrades 1.30\%, 1.45\%, and 1.82\% on pruning ResNet50, VGG16, and BERT respectively.
In contrast, the degradation of model accuracy of \sys is only 0.70\% and 0.85\% for ResNet50 and VGG on average.
\sys even improves the accuracy of the original model by 0.92\% on the CIFAR-100 dataset for ResNet50 and 0.24\% on the Moji dataset for BERT.


\noindent
{\bf Analysis}:
From~\autoref{tab:rq1-1}, we can have the following observations.
First, compared with other baselines, \sys can prune the given models and make them more effective (higher accuracy, precision, and recall) and fair (higher fairness performance).
Secondly, compared with the original models, the model pruned by \sys doesn't degrade the accuracy, and in some cases, the pruned models can achieve higher test accuracy.
This phenomenon could be attributed to the model pruning process enabling a concentrated emphasis on pivotal neurons, representing combinations of essential features, as noted in prior studies~\cite{wuFairPruneAchievingFairness2022,gaoFairNeuron2022}.
Therefore, in conclusion, \sys can effectively find and claim the winning ticket which can take both utility and fairness into account.
In other words, \sys can fix the bias issues in model pruning and retraining and guarantee the utility and fairness of the pruned model.

%% file: tftex/rq1_tab.tex
\begin{table}[]
    \caption{Effectiveness evaluation results. }\label{tab:rq1-1}
    \centering
    \footnotesize
    \tabcolsep=3pt
    \scalebox{0.85}{
    \begin{tabular}{cccrrrrr}
    \toprule
    Model & Dataset & Methods & \multicolumn{1}{c}{Acc.} & \multicolumn{1}{c}{Precis.} & \multicolumn{1}{c}{Recall} & \multicolumn{1}{c}{CWV} & \multicolumn{1}{c}{MCD} \\ \midrule
    \multirow{28}{*}{ResNet50} & \multirow{6}{*}{CIFAR-100} & Magnitude Pruning & 0.5892 & 0.5923 & 0.5881 & 0.0262 & 0.7493 \\
    & & Standard LTH & 0.6145 & 0.6186 & 0.6160 & 0.0254 & 0.7056 \\
    & & SafeCompress & 0.6163 & 0.6149 & 0.6168 & 0.0254 & 0.6396 \\
    & & FairScratch & 0.5957 & 0.5993 & 0.5952 & 0.0276 & 0.6407 \\
    & & \sys & {\bf 0.6449} & {\bf 0.6444} & {\bf 0.6436} & {\bf 0.0225} & {\bf 0.6243} \\ \cmidrule{3-8}
    & & Original Model & 0.6390 & 0.6417 & 0.6390 & 0.0229 & 0.6134 \\ \cmidrule{2-8}
    & \multirow{6}{*}{TinyImageNet} & Magnitude Pruning & 0.4136 & 0.4291 & 0.4296 & 0.1205 & 0.3565 \\
    & & Standard LTH & 0.4454 & 0.4268 & 0.4457 & 0.1193 & 0.3333 \\
    & & SafeCompress & 0.4460 & {\bf 0.4657} & 0.4436 & 0.1098 & 0.4242 \\ 
    & & FairScratch & 0.4133 & 0.4197 & 0.4153 & 0.1222 & 0.3436 \\ 
    & & \sys & {\bf 0.4534} & 0.4442 & {\bf 0.4512} & {\bf 0.1073} & {\bf 0.2500} \\ \cmidrule{3-8}
    & & Original Model & 0.4662 & 0.4377 & 0.4643 & 0.1173 & 0.2065 \\ \cmidrule{2-8}
    & \multirow{6}{*}{CelebA} & Magnitude Pruning & 0.9528 & 0.9501 & 0.9524 & 4.89e-4 & 0.0404 \\
    & & Standard LTH & 0.9691 & {\bf 0.9695} & 0.9687 & 3.87e-4 & 0.0384 \\
    & & SafeCompress & 0.9526 & 0.9619 & 0.9527 & 1.37e-4 & 0.0292 \\ 
    & & FairScratch & 0.9644 & 0.9649 & 0.9645 & 3.54e-4 & 0.0396 \\ 
    & & \sys & {\bf 0.9702} & 0.9689 & {\bf 0.9707} & {\bf 8.21e-5} & {\bf 0.0138} \\ \cmidrule{3-8}
    & & Original Model & 0.9705 & 0.9698 & 0.9711 & 3.13e-5 & 0.0109 \\ \cmidrule{2-8}
    & \multirow{6}{*}{LFW} & Magnitude Pruning & 0.9635 & 0.9628 & 0.9641 & 9.23e-4 & 0.0335 \\
    & & Standard LTH & 0.9700 & 0.9693 & 0.9706 & 7.30e-4 & 0.0315 \\
    & & SafeCompress & 0.9524 & 0.9517 & 0.9530 & 1.37e-4 & 0.0192 \\ 
    & & FairScratch & 0.9770 & 0.9763 & 0.9776 & 3.34e-4 & 0.0215 \\ 
    & & \sys & {\bf 0.9850} & 0.9843 & {\bf 0.9856} & {\bf 5.54e-5} & {\bf 0.0121} \\ \cmidrule{3-8}
    & & Original Model & 0.9865 & 0.9858 & 0.9871 & 3.59e-5 & 0.0123 \\ \midrule

    \multirow{28}{*}{VGG16} & \multirow{6}{*}{CIFAR-100} & Magnitude Pruning & 0.5899 & 0.5930 & 0.5888 & 0.0524 & 0.7979 \\
    & & Standard LTH & 0.6150 & 0.6191 & 0.6165 & 0.0508 & 0.7112 \\
    & & SafeCompress & 0.6168 & 0.6154 & 0.6173 & 0.0508 & 0.6784 \\
    & & FairScratch & 0.5962 & 0.5998 & 0.5957 & 0.0552 & 0.6821 \\
    & & \sys & {\bf 0.6456} & {\bf 0.6451} & {\bf 0.6443} & {\bf 0.0450} & {\bf 0.6486} \\ \cmidrule{3-8}
    & & Original Model & 0.6395 & 0.6422 & 0.6395 & 0.0458 & 0.6268 \\ \cmidrule{2-8}
    & \multirow{6}{*}{TinyImageNet} & Magnitude Pruning  & 0.4143 & 0.4298 & 0.4303 & 0.2410 & 0.3530 \\
    & & Standard LTH & 0.4461 & 0.4275 & 0.4464 & 0.2386 & 0.3669 \\
    & & SafeCompress & 0.4466 & {\bf 0.4663} & 0.4443 & 0.2196 & 0.4484 \\ 
    & & FairScratch & 0.4139 & 0.4203 & 0.4159 & 0.2444 & 0.3872 \\ 
    & & \sys & {\bf 0.4539} & 0.4448 & {\bf 0.4518} & {\bf 0.2146} & {\bf 0.2605} \\ \cmidrule{3-8}
    & & Original Model & 0.4667 & 0.4382 & 0.4648 & 0.2346 & 0.2130 \\ \cmidrule{2-8}
    & \multirow{6}{*}{CelebA} & Magnitude Pruning & 0.9533 & 0.9506 & 0.9530 & 0.0098 & 0.0308 \\
    & & Standard LTH & 0.9696 & {\bf 0.9700} & 0.9692 & 0.0077 & 0.0318 \\
    & & SafeCompress & 0.9531 & 0.9624 & 0.9532 & 0.0027 & 0.0304 \\ 
    & & FairScratch & 0.9649 & 0.9654 & 0.9650 & 0.0031 & 0.0292 \\ 
    & & \sys & {\bf 0.9707} & 0.9694 & {\bf 0.9712} & {\bf 0.0016} & {\bf 0.0276} \\ \cmidrule{3-8}
    & & Original Model & 0.9710 & 0.9703 & 0.9716 & 0.0063 & 0.0218 \\ \cmidrule{2-8}
    & \multirow{6}{*}{LFW} & Magnitude Pruning & 0.9642 & 0.9635 & 0.9648 & 6.30e-3 & 0.0470 \\
    & & Standard LTH & 0.9705 & 0.9698 & 0.9711 & 3.44e-3 & 0.0420 \\
    & & SafeCompress & 0.9529 & 0.9522 & 0.9535 & 5.61e-3 & 0.0264 \\ 
    & & FairScratch & 0.9775 & 0.9768 & 0.9781 & 3.32e-3 & 0.0230 \\ 
    & & \sys & {\bf 0.9855} & {\bf 0.9848} & {\bf 0.9861} & {\bf 1.61e-4} & {\bf 0.0233} \\ \cmidrule{3-8}
    & & Original Model & 0.9870 & 0.9863 & 0.9876 & 1.22e-4 & 0.0266 \\ \midrule
    
    \multirow{6}{*}{BERT} & \multirow{6}{*}{Moji} & Magnitude Pruning & 0.9188 & 0.9123 & 0.9142 & 0.4467 & 0.9984 \\
    & & Standard LTH & 0.9437 & 0.9472 & 0.9411 & 0.4469 & 0.9975 \\
    & & SafeCompress & 0.9336 & 0.9341 & 0.9318 & 0.3254 & 0.6966 \\
    & & FairScratch & 0.9430 & 0.9435 & 0.9412 & 0.2887 & 0.6536 \\
    & & \sys & {\bf 0.9636} & {\bf 0.9647} & {\bf 0.9566} & {\bf 0.1254} & {\bf 0.3323} \\ \cmidrule{3-8}
    & & Original Model & 0.9612 & 0.9523 & 0.9681 & 0.1168 & 0.3360 \\
    
    \bottomrule
    \end{tabular}
    }
    \vspace{-10pt}
\end{table}

%% file: body/rq2.tex
\subsection{RQ2: Efficiency in Finding Winning Ticket}\label{sec:rq2}

\input{tftex/rq2_tab.tex}


\noindent
{\bf Experiment Design}:
To evaluate the efficiency of \sys in finding and claiming the winning ticket, we measure the time cost of \sys and the baselines (i.e. Standard LTH, SafeCompress, and FairScratch) in performing a complete pruning and retraining process on these datasets.
Since the Magnitude Pruning method does not contain the retraining process after pruning, which makes it difficult to make a fair comparison, we do not involve this method in the comparison.
To avoid the effect of randomness, we perform 10 trials that use random training/test data splitting and compute and record the average time overhead of each method.
The corresponding results and analysis are presented below.

\noindent
{\bf Results}:
\autoref{fig:rq2} shows the time cost of each method and demonstrates the efficiency of \sys in finding fair sparse model.
The blue, orange, gray, and yellow bars represent the time cost of the four methods respectively.
On average, on the CIFAR-100, TinyImageNet, and CelebA datasets, \sys spends 255.8 minutes, 530.4 minutes, and 200.5 minutes, which are 16.80\%, 18.76\%, and 4.89\% faster than the SafeCompress, and 18.85\%, 12.05\%, and 5.07\% faster than the FairScratch.
Compared with the Standard LTH method, \sys takes slightly more time.
The average execution time of \sys is 3.02\%, 6.00\%, and 2.35\% longer than the Standard LTH.


\noindent
{\bf Analysis}:
As shown in~\autoref{fig:rq2}, compared to baselines, \sys costs close or less time on pruning and retraining.
\sys spent significantly less time than the SafeCompress method on finding and claiming lottery tickets, and its time cost is close to that of the Standard LTH.
Combined with the results in~\autoref{tab:rq1-1}, we can see that \sys achieves far better results than the baseline in terms of fairness and utility while costs similar or even less time, which demonstrates the efficiency of \sys in finding fair sparse models.

%% file: tftex/rq2_tab.tex



\begin{figure}[]
  \centering
  \scalebox{1.0}{
  \includegraphics[trim={0cm 3cm 0cm 1cm},clip,width=0.95\linewidth]{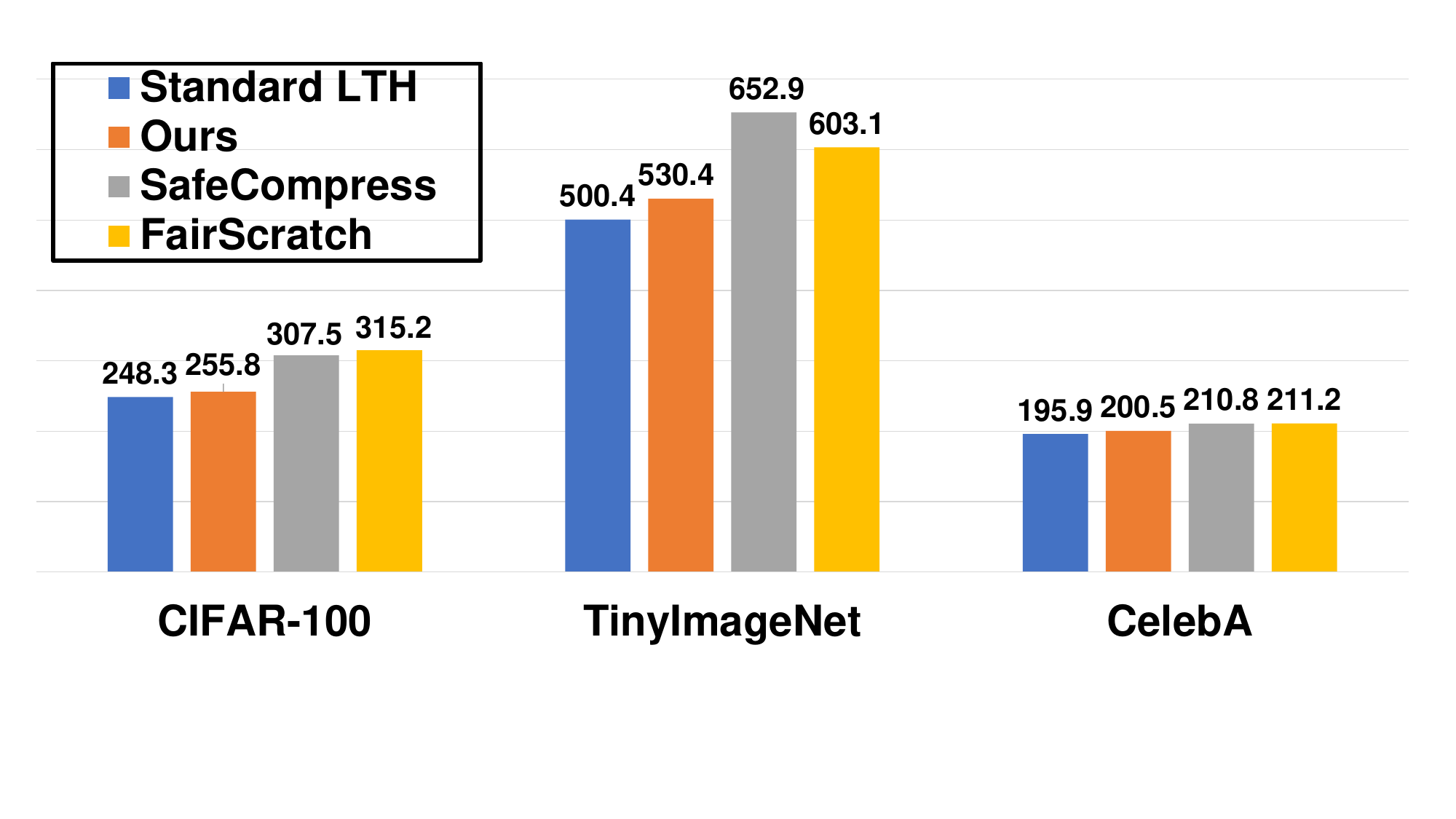}
  }
  \caption{Time to prune and retrain a model.}
  \label{fig:rq2}
  \vspace{-10pt}
\end{figure}



%% file: body/rq3.tex
\subsection{RQ3: Impacts of Mask Generation and Training Policies}
\label{sec:rq3}

\input{tftex/rq3_tab.tex}


\noindent
{\bf Experiment Design}:
\sys leverages two core components, namely the \textit{conflict-detection-based mask generation} and  \textit{training refinement}, to find and claim the winning tickets.
In this section, we aim to demonstrate the effectiveness of these two components by comparing the performance of the pruned model under different mask generation and training policies.
To this end, 1) for mask generation policies, we compare the results of finding winning tickets among the conflict-detection-based mask generation implemented in \sys (see~\autoref{sec:mask}), random mask generation, and magnitude mask generation.
The random policy randomly selects neurons to generate masks.
The magnitude one removes those neurons with the smallest weights to generate a mask, which is applied in Standard LTH~\cite{frankle2018lottery}.
2) for training policies, we compare the performance among \sys rewinding to 5th epoch, \sys rewinding to 10th epoch~(default), \sys rewinding to 15th epoch, \sys without learning rate decreasing, and \sys without rewind.
The latter two are training policies that do not use learning rate decreasing or the rewind method, which are introduced in~\autoref{sec:training}.
We follow the experiment settings in~\autoref{sec:setup} and conduct experiments on the CIFAR-100 dataset.
We record and compare the accuracy, precision, recall, CWV, and MCD of these models to observe how different mask and training policies affect their utility and fairness.

\noindent
{\bf Results}:
\autoref{tab:rq3-1} presents the results, where the best results are marked in bold.
The first columns list different mask generation and training policies, and the following five columns show the results of the pruned models on accuracy, precision, recall, CWV, and MCD.
Overall,  \sys achieve the comprehensively best fix results among the compared methods, which indicates that the conflict-detection-based mask generation and training refinement in \sys can help to find fair sparse models.

\textit{Impact of mask generation policy:}
The first three rows of~\autoref{tab:rq3-1} show the comparison results of mask generation methods, where the first line shows the results of the default method in \sys.
When \sys applies the random mask generation instead of conflict-detection-based mask generation, the accuracy of pruned models decreases by 1.02\%, and CWV and MCD increase by 15.87\% and 4.73\%.
In addition, using the magnitude mask generation method instead of the one in \sys also causes performance degradation.
It decreases the accuracy by 0.74\% and increases CWV and MCD by 14.67\% and 7.51\% respectively.
It follows that the model obtained by BALLOT pruning is more effective in maintaining fairness, in line with our observations in \autoref{sec:ticketselection}.

\textit{Impact of training policy:} The last five rows show the effect of different training policies.
Compared to the default training refinement in \sys, \sys without learning rate decreasing and \sys without rewind reduce the accuracy of pruned models by 8.17\% and 2.54\% and increase the CWV by 25.47\% and 4.44\%.
We can observe that removing any method in the learning rate decreasing and the rewind will cause a significant degradation in the performance of \sys.
In the context of employing the rewind method, loading the weights from the 10th epoch yields superior performance compared to loading from the 5th and 15th epoch, resulting in an increase of 1.75\% and 3.43\% in CWV and 1.51\% and 0.92\% in MCD, respectively.

\noindent
{\bf Analysis}:
The above results show that compared with the other optional methods, the conflict-detection-based mask generation and training refinement in \sys achieves the highest utility and fairness performance, which can improve model fairness when ensuring effectiveness.
In addition, both mask generation and training policy implemented in \sys is helpful to improve the performance of the pruned model.
It is noteworthy that \sys rewidning to 10th epoch exhibits marginal superiority over both 5th epoch and 15th epoch. 
This observation aligns with the theoretical framework proposed in~\cite{frankle2020linear}, suggesting that the original network's weights are prone to stability against stochastic gradient descent~(SGD) noise after an initial training period. 
However, an extended training duration may result in weight solidification.

%% file: tftex/rq3_tab.tex
\begin{table}[]
    \caption{Comparison among different mask generation and training policies in \sys. \textit{w/o} is short for \textit{without.}, and \sys~($n$th) is short for \sys rewinding to $n$th epoch.}\label{tab:rq3-1}
    \centering
    \footnotesize
    \tabcolsep=3.5pt
    \scalebox{0.9}{
    \begin{tabular}{ccrrrrr}
    \toprule
    \multicolumn{2}{c}{Method} & \multicolumn{1}{c}{Acc.} & \multicolumn{1}{c}{Precis.} & \multicolumn{1}{c}{Recall} & \multicolumn{1}{c}{CWV} & \multicolumn{1}{c}{MCD} \\ \midrule
    \multirow{3}{*}{Mask Policy} & Random & 0.6383 & 0.6353 & 0.6379 & 0.0261 & 0.6538 \\
    & Magnitude & 0.6401 & 0.6395 & 0.6410 & 0.0258 & 0.6712 \\
    & \sys & {\bf 0.6449} & {\bf 0.6444} & {\bf 0.6436} & {\bf 0.0225} & {\bf 0.6243} \\ \midrule
    \multirow{5}{*}{Training Policy} & w/o LR decreasing & 0.5922 & 0.5905 & 0.5924 & 0.0282 & 0.7512 \\
    & w/o rewind & 0.6285 & 0.6237 & 0.6277 & 0.0235 & 0.6566 \\
    & \sys~(5th) & 0.6411 & {\bf 0.6448} & 0.6423 & 0.0229 & 0.6339 \\
    & \sys~(10th) & {\bf 0.6449} & 0.6444 & {\bf 0.6436} & {\bf 0.0225} & {\bf 0.6243} \\
    & \sys~(15th) & 0.6425 & 0.6421 & 0.6431 & 0.0233 & 0.6301 \\
    \bottomrule
    \vspace{-14pt}
    \end{tabular}
    }
\end{table}

%% file: body/rq4.tex
\subsection{RQ4: Impacts of Configurable Parameters}
\label{sec:rq4}


\input{tftex/rq4_fig.tex}

\noindent
{\bf Experiment Design}:
\sys leverages two hyperparameters \(\gamma\) and \(\eta\) in mask generation.
The former is used to control the contribution of fairness loss \(L_f\) and accuracy loss \(L_a\).
The latter controls the neuron sorting ratio, which determines which neurons in each iteration will be considered to have the highest degree of conflict and increment the count.
We conduct experiments to investigate and understand how different values of these configurable hyperparameters affect the performance of \sys in finding and claiming the winning ticket.
In the experiment, we run \sys on the CIFAR-100 dataset with different \(\log{\gamma}\) values from -1.0 to 2.0 and different \(\eta\) values from 0.7 to 0.99.
To evaluate the effect of different values of configurable parameters, we record the accuracy, CWV, and MCD of the model pruned by \sys with different parameter settings.

\noindent
{\bf Results}:
\autoref{fig:rq4} shows the comparison results of different values of the configurable parameters \(\gamma\) and \(\eta\),
where~\autoref{fig:rq4_1} and~\autoref{fig:rq4_3} presents how \(\gamma\) and \(\eta\) affect the accuracy of the pruned models in \sys, and~\autoref{fig:rq4_2} and~\autoref{fig:rq4_4} show the effect of different values of \(\gamma\) and \(\eta\) on the fairness of the pruned model (i.e., CWV and MCD).

\textit{Impact of \(\gamma\):}
\sys uses the parameter \(\gamma\) to control the impact of the fairness loss in mask generation.
A higher \(\gamma\) leads the pruned model to pay more attention to fairness.
As shown in~\autoref{fig:rq4_1} and~\autoref{fig:rq4_2}, when the value of \(\log{\gamma}\) increases, the accuracy of the model pruned by \sys first increases and then decreases, while the values of CWV and MCD generally show a downward trend.
When \(\log{\gamma}\) is 1.0, the accuracy of the pruned models reaches the highest value of 64.49\%.
It is noteworthy that at the two points of 0.5 and 1.0, there are abnormal increases in the values of CWV and MCD, and then their values continue to decline as the value of \(\log{\gamma}\) increases.
However, the increment brought by these outliers is relatively small, and the impact on the model fairness is not significant.
Considering the trade-off between accuracy and fairness, \sys selects 1.0 as the default value of \(\log{\gamma}\) (i.e., \(\gamma\) is set to 10) to maximize the effectiveness on improve model fairness and utility.

\textit{Impact of \(\eta\):}
\(\eta\) is used to determine which neurons will be regarded as conflicting neurons and counted.
A higher \(\eta\) indicates that in mask generation of \sys, the criteria for evaluating conflicting neurons are more strict.
The experiment results in~\autoref{fig:rq4_3} and~\autoref{fig:rq4_4} show that the model performance of both utility and fairness first increase and then decreases as \(\eta\) rises, and \sys performs best when \(\eta\) is set to 0.95.
Therefore, we choose 0.95 as the default value of \(\eta\).

\noindent
{\bf Analysis}:
From~\autoref{fig:rq4}, we can observe that different values of \(\gamma\) and \(\eta\) have a significant impact on the performance of the pruned model.
In terms of accuracy, as the values of \(\gamma\) and \(\eta\) increase, the model accuracy first increases and then decreases.
Similarly, in terms of fairness, the increase of \(\eta\) first improves the performance of the model fairness and then worsens it.
In addition, the increase of \(\gamma\) improves the model fairness in general.
As a result, to effectively select the winning ticket and prune a sparse model toward both accuracy and fairness, we set the default value of \(\gamma\) to 10, and the value of \(\eta\) to 0.95 in \sys.

\noindent
\textbf{Summarization}.
We have proved the advancement of BALLOT through the above experimental evaluations.
Compared to the baseline, BALLOT effectively balances fairness and accuracy throughout the pruning process and efficiently identifies equitable subnetworks, thereby expediting training. 
BALLOT requires a more intricate procedural approach than conventional pruning methods, resulting in higher training costs than direct pruning. However, this additional expense remains relatively modest, averaging no more than a 6\% increase.

%% file: tftex/rq4_fig.tex
    

\begin{figure}
    \centering     
    \subfloat[\(\gamma\)-accuracy]
    {\includegraphics[width=0.22\textwidth,height=0.18\textwidth]{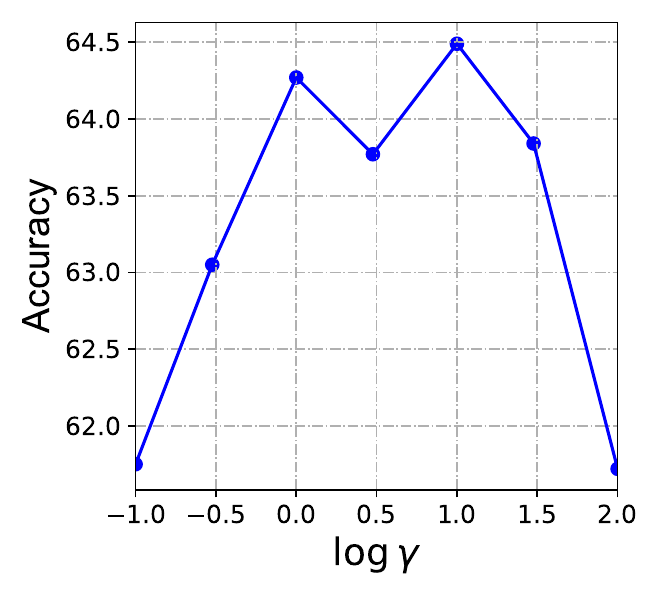}
    \label{fig:rq4_1}}
    \subfloat[\(\gamma\)-CWV\&MCD]
    {\includegraphics[width=0.22\textwidth,height=0.18\textwidth]{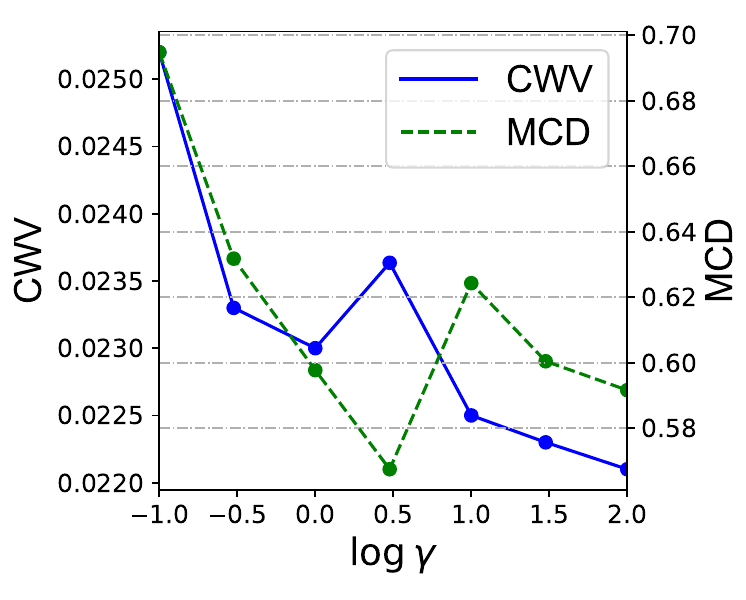}
    \label{fig:rq4_2}}
    
    \subfloat[\(\eta\)-accuracy]
    {\includegraphics[width=0.22\textwidth,height=0.18\textwidth]{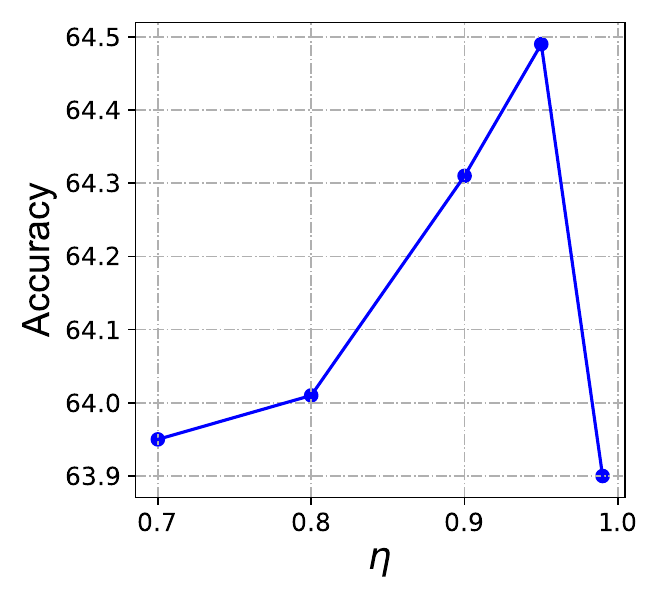}
    \label{fig:rq4_3}}
    \subfloat[\(\eta\)-CWV\&MCD]
    {\includegraphics[width=0.22\textwidth,height=0.18\textwidth]{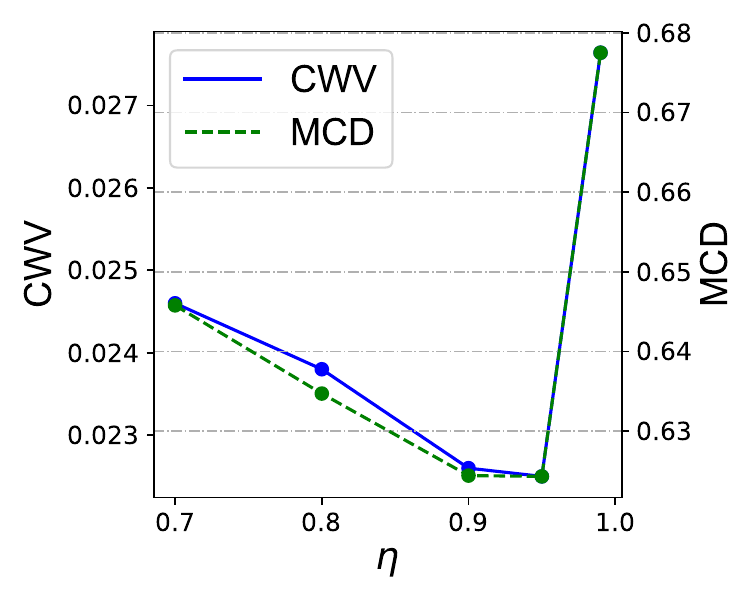}
    \label{fig:rq4_4}}
  \caption{Effect of \(\gamma\) and \(\eta\) on model average performance.}
  \label{fig:rq4}
  \vspace{-17pt}
\end{figure}

%% file: body/RelatedWork.tex
\section{Threat to Validity}\label{sec:threat}

\sys is evaluated on five mainstream datasets with ResNet50, VGG16, and BERT under the sparsity of 0.05, which may be limited. 
Additionally, the existence of several configurable parameters introduces some threats.
While our experiments demonstrated promising pruning results, the effectiveness remains uncertain and challenging when applied to more complex and advanced models like transformer.
To address those concerns, we have taken the step of open-sourcing the implementation of \sys and providing comprehensive experimental details that encompass models before and after pruning, as well as training configurations.
For reproduction purposes, all codes and data are available at~\cite{AnonymizedRepository}.

\vspace{-10pt}

\section{Related work and discussion}\label{sec:rw}

\noindent\textbf{Model Compression.}
Model compression aims to reduce the size and computational complexity of large neural network models. 
There are various model compression methods, such as network pruning, parameter quantization, and knowledge distillation.

\textit{Network pruning.} 
Pruning methods focus on removing redundant components without sacrificing performance significantly.
Channel-based pruning removes entire channels (feature maps) from convolutional neural networks~\cite{polyak2015channel}.
Filter-based pruning selectively removes individual filters (kernels)~\cite{he2019filter,li2016pruning}.
The unimportant connections selection is based on criteria such as weight magnitude~\cite{han2015deep,han2015learning}, gradients~\cite{molchanov2019importance}, and hessian~\cite{lecun1989optimal} statistics.

\textit{Parameter quantization.} 
Parameter quantization of neural networks is the process of converting the weights and
activation values of a network model from high precision to low.
Quantization-aware training~(QAT)~\cite{ni2020wrapnet} introduces a pseudo-quantization method to simulate the error brought by the quantization process.
Post-Training quantization~(PTQ) methods~\cite{jacob2018quantization} directly quantize the pre-trained model without retraining, which is more efficient but may cause more accuracy loss.

\textit{Knowledge distillation.}
Knowledge distillation is a training paradigm with teacher-student architecture.
The teacher network~(a complex pre-trained network) provides the student network~(a simple small network) with prior knowledge so that the student network achieves similar performance to that of the teacher.
Researchers have tried to train student networks using different prior knowledge~\cite{hinton2015distilling,bengio2013representation}.

The software engineering community has exhibited significant interest in model compression. 
Shi et al. investigate the application of knowledge distillation techniques to condense pre-trained code models, reducing their size to a mere 3MB, thereby rendering them easily deployable~\cite{shi2022compressing}. 
Zhu et al. delve into the advantages of employing test-driven development for pruning, incorporating bi-objective optimization to balance performance and safety properties~\cite{zhu2022safety}.
Quality assurance for the deployment of compression-based AI models is a subject of great concern. 
Xie et al. conduct differential testing to compare the behaviors of models before and after compression~\cite{xie2019diffchaser}. 
Additionally, Zhang et al. propose an ILP-based formal verification approach for quantized neural networks~\cite{zhang2022qvip}.

\noindent\textbf{Fairness of ML.}
The issue of fairness in ML is gaining increasing prominence, in the areas of CV~\cite{tian2022image}, NLP~\cite{yang2021biasrv,asyrofi2021biasfinder} and especially critical automated decision-making systems like higher education~\cite{clearyTestBiasValidity1966}, employment~\cite{guionEmploymentTestsDiscriminatory1966, raghavanMitigatingBiasAlgorithmic2020, vandenbroekHiringAlgorithmsEthnography2019}, and re-offense judgement~\cite{oneilWeaponsMathDestruction2016, brennanEmergenceMachineLearning2013}.

To address the concern above, fairness testing for ML systems is gaining attention in the software engineering community.
AEQUITAS proposes a directed search for individual discriminatory instances ~\cite{udeshiAutomatedDirectedFairness2018}.
Symbolic Generation~(SG) integrates symbolic execution and local model explanation to craft individual discriminatory instances~\cite{agarwalAutomatedTestGeneration2018}.
ExpGA uses the genetic algorithm to generate discriminatory instances~\cite{fan2022explanation}.
Besides, ADF and EIDIG combine global search and local search to systematically explore the input space with the guidance of gradient~\cite{zhangWhiteboxFairnessTesting2020, EIDIG2021efficient}.
There are also concerns about group fairness.
THEMIS considers group fairness using causal analysis and uses random test generation to evaluate fairness~\cite{angellThemisAutomaticallyTesting2018}.
A recent framework namely FairRec is dedicated to uncovering disadvantaged groups in recommender systems~\cite{guo2023fairrec}.

While uncovering fairness issues through testing, researchers attempt to mitigate and repair fairness deficiencies.
Pre-processing approaches focus on mitigate dataset bias by correcting labels~\cite{kamiranClassifyingDiscriminating2009, zhangAchievingNonDiscriminationData2017}, revising attributes~\cite{feldmanCertifyingRemovingDisparate2015, kamiranDataPreprocessingTechniques2012}, generating non-discrimination data~\cite{sattigeriFairnessGANGenerating2019, xuFairganFairnessawareGenerative2018}, and constructing fair data representations~\cite{beutelDataDecisionsTheoretical2017,liTrainingDataDebugging2022}.
In-processing and post-processing approaches aim to mitigate algorithm bias by improving the learning process or the learned model.
More specifically, these approaches propose an objective function considering the fairness metric of prediction~\cite{zhangMitigatingUnwantedBiases2018}, adapt fairness-driven generative adversarial framework~\cite{xuFairganFairnessawareGenerative2018,adelOneNetworkAdversarialFairness2019,gaoFairneuronImprovingDeep2022}, or directly change the predictive labels of bias models' output~\cite{hardtEqualityOpportunitySupervised2016, pleissFairnessCalibration}, etc.
In the SE community, the most common way to improve model fairness is to retrain based on testing-generated discriminatory instances~\cite{zhangWhiteboxFairnessTesting2020, EIDIG2021efficient, fan2022explanation}. 
Further, Sun et al. propose to apply causal analysis to identify neurons that are guilty of introducing bias~\cite{sun2022causality}.
Gao et al. adapt path analysis to selecting neurons that lead to conflicting accuracy and fairness joint-optimizations~\cite{gaoFairNeuron2022}.
Recently, the fairness issue in the maintenance and deployment of ML models has arrived at concern, e.g. CILIATE is proposed to mitigate bias amplification in the incremental learning process~\cite{gao2023fairCILciliate}.

\noindent\textbf{Discussion.}
To the best of our knowledge, none of the current repairing methods have been designed to consider addressing the fairness issues of model compression typified by pruning.
We aim to raise awareness of fairness issues in model compression, which is a critical step towards the development of responsible AI systems.


%% file: body/conclusion.tex
\section{Conclusion}\label{s:conclusion}

Inspired by ethics-aware software engineering, we propose and develop \sys, an innovative fairness-aware deep neural network pruning framework, powered by conflict-detection-based mask generation and training refinement.
It can identify accurate and fair tickets~(pruned model prototype) and refine the ticket training process to obtain the high-performance pruned models.
Our evaluation results show that \sys effectively mitigates pruning fairness problems and outperforms all baselines in terms of fairness and accuracy.

\section{Data Availability}\label{s:data}
All source code and data used in our work can be found at~\cite{AnonymizedRepository}.